\documentclass{article}
\usepackage[latin1]{inputenc} 
\usepackage{amssymb}

\newcommand{\La}{\langle}
\newcommand{\Ra}{\rangle}
\newcommand{\ep}{\langle\rangle}
\newcommand{\Let}{\mathsf{let}}
\newcommand{\In}{\mathsf{in}}

\newcommand{\If}{\mathsf{if}}
\newcommand{\Then}{\mathsf{then}}
\newcommand{\Else}{\mathsf{else}}

\newcommand{\True}{\mathsf{true}}
\newcommand{\False}{\mathsf{false}}
\newcommand{\Bool}{\mathsf{bool}}
\newcommand{\Int}{\mathsf{int}}
\newcommand{\List}{\mathsf{list}}

\newcommand{\Case}{\mathsf{case}}
\newcommand{\Of}{\mathsf{of}}

\newcommand{\Do}{\mathsf{do}}

\newcommand{\q}{\mathsf{q}}

\newcommand{\ql}{\mathsf{li}}
\newcommand{\qu}{\mathsf{un}}

\newcommand{\gl}{\mathsf{gl}}
\newcommand{\lo}{\mathsf{lo}}
\newcommand{\qh}{\mathsf{hi}}

\newcommand{\sq}{\mathsf{\varrho}}
\newcommand{\bt}{\mathsf{t}}
\newcommand{\va}{\mathsf{v}}
\newcommand{\op}{\mathsf{o}}
\newcommand{\e}{\mathsf{e}}

\newcommand{\p}{\mathsf{p}}
\newcommand{\pp}{\overline{\mathsf{p}}}

\newcommand{\x}{\mathsf{x}}
\newcommand{\y}{\mathsf{y}}
\newcommand{\z}{\mathsf{z}}
\newcommand{\w}{\mathsf{w}}
\newcommand{\n}{\mathsf{n}}
\newcommand{\f}{\mathsf{f}}
\newcommand{\g}{\mathsf{g}}
\newcommand{\G}{\mathsf{g}}
\newcommand{\arb}{\mathsf{b}}
\newcommand{\arc}{\mathsf{c}}

\newcommand{\Fun}{\mathsf{f}}

\newcommand{\xs}{\mathsf{xs}}

\newcommand{\id}{\mathsf{id}}

\newcommand{\E}{\mathsf{E}}
\newcommand{\T}{\mathsf{T}}

\newcommand{\Pt}{\mathsf{P}}

\newcommand{\V}{\mathsf{V}}
\newcommand{\B}{\mathsf{B}}

\newcommand{\Array}{\mathsf{array}}

\newcommand{\ara}{\mathsf{a}}

\newcommand{\Ins}{\mathsf{ins}}

\newcommand{\Map}{\mathsf{map}}
\newcommand{\Fib}{\mathsf{f}}
\newcommand{\Fact}{\mathsf{f}}

\newcommand{\vi}{\mathsf{i}}

\newcommand{\F}{\mathsf{F}}
\newcommand{\St}{\mathsf{S}}
\newcommand{\cf}{\mathsf{c}}
\newcommand{\zs}{\mathsf{zs}}
\newcommand{\ct}{\mathsf{c}}

\begin{document}
\begin{center}
\textbf{\large{Weak-linearity, globality and in-place update}}\medskip\\
Héctor Gramaglia\medskip\\
FAMAF, Facultad de Matemática, Astronomía, Física y Computación\\
Universidad Nacional de Córdoba\\
CIEM, Centro de Investigación y Estudios de Matemática
\end{center}
\noindent \textbf{Abstract: }
\textit{
Computational interpretations of linear logic allow static control of memory resources: the data produced by the program are endowed through its type with attributes that determine its life cycle. This has promoted numerous investigations into safe introduction of in-place update. Various type systems have been proposed for this aim, but the memory management that promotes linear evaluation does not adequately model the destruction of in-place update.
The main achievement of this work is to establish a simple theoretical framework that will allow us to clarify the potential (and limits) of linearity to guarantee the process of transforming a functional program into an imperative one.
For this purpose we will introduce a type system called \textit{global} that will model the in-place update as the linear system models the one-time use.
}

\section{Introduction}
\label{introduccion}

In the formulae-as-types interpretation of Girard's linear logic \cite{girard}, the type of a value is not only a description of its "form", but also, in its  computational interpretation, an ability to use it a certain number of times. This refinement plays a key role in advanced type systems that are developed for a variety of purposes, including static resource management and concurrent systems. In particular, much research focuses on obtaining imperative implementations of functional programs that, by modifying data in-place, provide an efficient alternative to the traditional implementation, in which garbage collection is necessary to rescue heap space. But the linearity property is too restrictive in practice, more than necessary to guarantee the correctness of the in-place update. Many works address the problem of weakening the notion of linearity for different specific purposes (Wadler \cite{wadler}, Odersky \cite{odersky}, Kobayashi \cite{kobayashi}, Smith, Walker, and Morrisett \cite{smith}, Aspinall and Hofmann \cite{aspinall}, Foster, Terauchi, and Aiken \cite{foster}, Aiken, Foster, Kodumal, and Terauchi \cite{aiken}, Gramaglia \cite{gramagliawlt}).

The main objective of this work is to clarify the potential of the process of attributing (weak) linearity to the data of a functional program as a safe way to introduce in-place update and global variables. Although this process has been addressed in many works, it is not easy to establish how successful it is. 
We will do this by introducing a type system called \textit{global}, which will express in types the information about the memory destruction caused by update-in-place, in the same way that linear types express information about the memory destruction produced by discarding a data after its only use.
In this way, the two forms of destruction can be properly compared, and it can be easily defined when the linearity property guarantees safe in-place update.

We are going to introduce two useful concepts: \textit{linear ratio} and \textit{functional residue}.
The linear ratio quantifies the improvement in the use of memory resources of the linear program in relation to the functional one. The functional residue quantifies the part of the improvement of the linear version that cannot be capitalized by the imperative program\footnote{From another point of view, it quantifies the failure of linearity when trying to model the in-place update.}.

More precisely, in \cite{gramagliawlt} a linear language is presented that uses a \textit{qualified signature} $\Sigma^\q$ to provide weak-linear\footnote{In this work a third qualifier $\qh$ (hiding) is used  to relax the linearity property.}   attributes to the program data. We will use a modification of this language that limits the presence of relevant qualifiers to $\Sigma^\q$. Although this decision implies a loss of generality, it is crucial: by limiting the presence of qualifiers, we will be able to compare the level of destructiveness of different forms of evaluation.

The strategy of condensing the linear attributes in the signature $\Sigma^\q$ will be complemented by the strategy of condensing the imperative aspects of a program in another qualified signature $\Sigma^\g$ (\textit{global signature}). Correct linear typing (using the $\Sigma^\q$ operators), together with correct global typing (using the $\Sigma^\g$ operators), added to a compatibility relationship $(\vartriangleright)\subseteq\Sigma^\q\times \Sigma^\g$, will grant ``protection'': the linearity property will be a guarantee of correctness for the introduction of in-place update and global variables.

The signatures $\Sigma^\qu$, $\Sigma^\q$ and $\Sigma^\g$ completely determine the three versions of a program, respectively, the functional one (all qualifiers are unrestricted), the linear one and the imperative. The linear ratio will measure the improvement of the linear operators $\Sigma^\q$ with respect to the unrestricted $\Sigma^\qu$, and the functional residual represents the level of inability of $\Sigma^\g$ to reproduce the improvement achieved by $\Sigma^\q$.


For a complete description of the history of substructural logics and their applications to Computer Science see \cite{walker} and \cite{dosen}. Several works use ideas similar to the qualifier $\qh$ of \cite{gramagliawlt}. We can mention in this line Wadler's \textit{sequential let}  \cite{wadler}, the \textit{usage aspect} given by Aspinally Hofmann in \cite{aspinall}, the \textit{observer annotations} of Oderskyn in  \cite{odersky}, and the  \textit{quasi linear types} of Kobayashi in  \cite{kobayashi}. The distinctive character of \cite{gramagliawlt} is that the main virtue of the formulation given by Walker in \cite{walker} is preserved: substructurality is completely captured by the introduction of context splitting, as the only modification to a classical type system.

Various works define translations from functional programs to imperatives, guaranteeing correctness through static analysis, in many cases based on variants of linear logic to control the single-threadedness property: Aspinall and Hofmann \cite{aspinall},   Chirimar, Gunter and  Riecke \cite{chirimar}, Draghicescu and Purushothaman \cite{draghicescu}, Hofmann \cite{hofmann}, Kobayashi \cite{kobayashi}, Shankar  \cite{shankar}, Wadler \cite{wadler}. A work that focuses exclusively on the introduction of global variables is Sestoft \cite{sestoft}.

Our approach aims to establish a simple theoretical framework to go beyond a correct translation of an applicative program into an imperative: to outline a theoretical framework that allows addressing the complex problem of the coexistence of both paradigms.

\section{Weak-linear programs}
\label{unlenguajeaplicativo}

Our linear language is built from a qualified heterogeneous signature $\Sigma^\q$, which is defined in Figure 1 from a heterogeneous signature $\Sigma$. 
For notational convenience we include list constructors within  $\Sigma^\q$.

\begin{center}
\begin{tabular}{|c |}
\hline
$
\begin{array}{llll}
\q & ::= & \ql\ |\ \qu\ & \mathsf{qualifier}\smallskip\\
\varrho & ::= & \ql\ |\ \qu\ |\ \qh& \mathsf{pseudoqualifier}\smallskip\\
\B & ::= & \Int\ |\ \Bool\ |\  \Array\ |\ ...&  \mathsf{basic\ pretype}\smallskip\\
\Pt& ::= & \B\ |\ [\E]  &\mathsf{storable\ pretype}\smallskip\\
\E& ::= & \q\ \Pt &\mathsf{storable\  type}\smallskip\\
E& ::= & \sq\ \Pt &\mathsf{storable\ pseudotype}\smallskip\\
\sigma & ::= & \q\ [\E]\ |\ (\E,\q\ [\E])\rightarrow\q\ [\E]&\mathsf{constructor \ types}\smallskip\\
\tau & ::= & (\sq_1\ \B_1,...,\sq_n\ \B_n)\rightarrow\q\ \B\ &\mathsf{operator \ type}\smallskip\\
\Sigma^\q & ::=  &\{(\op^\tau:\tau)\ :\ (\op:(\B_1,..., \B_n)\rightarrow\B)\in \Sigma\}\ \cup\ \   &\mathsf{qualified \ signature}\\
 &  &\{([]^\sigma:\sigma)\}\cup\{(:)^\sigma: \sigma\} \\
\end{array}
$\\
\hline
\end{tabular}
\tiny Figure 1: Qualifiers, storable types and qualified signature\end{center}

Qualifying the basic types will allow us to obtain different forms of evaluation for our language. Roughly speaking, we have three modalities for a storable pseudotype $\sq\ \Pt$
(the $\qh$ qualifier will only be used for storable pseudotypes in the role of input).
The \textit{unrestricted} mode, represented by $\qu\ \Pt$, indicates that the data can be used an unlimited number of times.
The \textit{linear} mode ($\ql\ \Pt$) indicates that the data will be used once (without being hidden), and the \textit{hidden} mode ($\qh\ \Pt$), indicates read-only use of a linear data (it is not deallocated from memory).

The abstract syntax of the language $L^1[\Sigma^\q]$ is shown in Figure 2. The abstract phrase $\x$ represents an infinite set of variables. By $\op$ (without arguments) we denote the constants of $\Sigma$, that is, the function symbols of arity 0.


The main differences between $L^1[\Sigma^\q]$ and the language presented in \cite{gramagliawlt} are that the former limits the presence of qualifiers to the pseudotype of operators and constructors\footnote{Note that the values in $\St$ do not have qualifiers.} and the phrase $\Case$. Neither tuples nor lambda terms are preceded by a qualifier. We will see in the next section that tuples will not be values that are stored in memory, and lambda terms will always be interpreted as unrestricted. The fact that tuples are not storable values  forces us to generalize the lambda abstraction allowing patterns. This, and the other modifications that $L^1[\Sigma^\q]$ presents with respect to the language presented in \cite{gramagliawlt}, aim to address the problem of the introduction of in-place update and global variables.

The phrase $\Let\ \p\equiv \e\ \In\ \e'$ replaces the split of \cite{walker}, a necessary construction in substructural systems due to the restriction on the number of uses of program objects. Since no tuples will be stored in memory, the phrase $\Let$ has the meaning of a local definition.

As will be seen in section \ref{semanticassmallstep}, function definitions in $\St$ will be considered recursive as long as the variable being defined is free in the body of the definition.

\begin{center}
\begin{tabular}{|c |}
\hline
$
\begin{array}{ll}
\begin{array}{lll}
\p & ::= & \x \\
&  & \La \p_1,...,\p_n \Ra \medskip\\
\va & ::= & \op\quad\ \ (\op\in\Sigma\textsf{ of arity 0})\\
    &     & []\\
    &     & (\x_1\!:\x_2)\\
    &     & \lambda\p.\e\smallskip\\
\St & ::= & \emptyset\\
    &     & \x=\va,\ \St\\
\end{array}
&
\begin{array}{rlllll}
\e& ::= & \x                    \\
  &     & \op^\tau( \e_1,...,\e_n)     \\
  &     & []^\sigma \\
  &     & ( \e_1:\e_2)^\sigma   \\
  &     & \La \e_1,...,\e_n \Ra \\
  &     & \x\ \e \\
   &    & \Let\ \p\equiv \e_1\ \In\ \e_2 \\
  &     & \If\ \e_1\ \Then\  \e_2\ \Else\ \e_3 \\
  &    &  \Case^\q\ \e_1\ \Of\ (\e_2,(\z_1\!\!:\!\z_2)\!\!\rightarrow\!\e_3)
\end{array}
\end{array}
$\\
\hline
\end{tabular}
\tiny Figure 2: Syntax of $L^1[\Sigma^\q]$
\end{center}

A program of $L^1[\Sigma^\q]$ will be a pair of the form $(\St,\e)$. In  section \ref{glosariodecasosdeestudio} numerous examples can be found, which will be used to study linearizations and globalizations.

\subsection{A linear system for $L^1[\Sigma^ \q]$}
\label{unsistemaslinealparal1sigma}

The linear system that we present below has its origin in the system defined in \cite{gramagliawlt}. Types, pseudotypes and type context are defined in Figure 3.

\begin{center}
\begin{tabular}{|c |}
\hline
$
\begin{array}{ll}
\begin{array}{llll}
\T  ::= & \E \qquad\qquad \mathsf{Expression\  type}\\
    & \T_1\rightarrow\T_2\\
      & \La\T_1,...,\T_n\Ra\smallskip\\
\end{array}
&
\begin{array}{rlllll}
\\
\V  ::= & E  & \mathsf{Value\  pseudotype}\\
&   \T_1\rightarrow\T_2\smallskip\\
\Gamma     ::=    & []\  & \mathsf{Type\  context}\\
 & \Gamma,\ \x\ :\V\\
\end{array}
\end{array}
$\\
\hline
\end{tabular}
\tiny Figure 3: Value types, expression types and type contexts
\end{center}

As usual, we allow a given variable to appear at most once in a context.

To preserve one of the invariants of linear systems we need to garantee that unrestricted data structures do not hold objects with linear types. To check this,  we define the predicate $\q(\V)$ by the following condition: $\q(\T\rightarrow\T')=\True$ and $\q(\sq\ \Pt)=\True   $ if and only if $\q=\ql$ o $\sq\neq\ql$.  The extension $\q(\Gamma)$ of predicate $\q(\V)$ to type contexts is immediate. We denote by $\Gamma^\qu$ the largest subcontext of $\Gamma$ that satisfies $\qu(\Gamma)$.

A central device of this system is the \textit{context split} $\Gamma_1\circ...\circ \Gamma_n=\Gamma$, a $(n+1)$-ary relation defined in Figure 4. For simplicity we will define the split for $n=2 $. The reader will have no difficulty in obtaining the definition for the general case.

\begin{center}
\begin{tabular}{|c |}
\hline
$
\begin{array}{cc}
\begin{array}{c}
\\
\overline{[] \circ []\ =\ []}\\
\end{array}&
\begin{array}{c}
\\
\Gamma_1\circ\Gamma_2=\Gamma\qquad(\V\neq\ql\ \Pt)\\
\overline{(\Gamma_1,\x: \V) \circ(\Gamma_2,\x:  \V)=\Gamma,\x:  \V}\\
\end{array}
\\
\\
\begin{array}{c}
\Gamma_1\circ\Gamma_2=\Gamma\\
\overline{(\Gamma_1,\x: \ql\ \Pt) \circ\Gamma_2\ =\ \Gamma,\x:  \ql\ \Pt\qquad}\\
\\
\end{array}&
\begin{array}{c}
\Gamma_1\circ\Gamma_2=\Gamma\\
\overline{\Gamma_1 \circ(\Gamma_2,\x: \ql\ \Pt)\ =\ \Gamma,\x:  \ql\ \Pt}\\
\\
\end{array}

\end{array}
$\\
\hline
\end{tabular}
\tiny Figure 4: Context split
\end{center}

\noindent For convenience, we define the $(0+1)$-ary case as $\qu(\Gamma)$\footnote{It is relevant in the rule for $\op^\tau$, when $\op$ is a constant symbol, that is  $\tau=\q\ \B $ (Figure 5).}.

But the context split, which is suitable for the typing of terms, is not suitable for the typing of expressions in general. By typing these, we must generate the possibility of a hidden use of a data as input of a basic operation.
For this, in \cite{gramagliawlt} we define the \textit{context pseudosplit}. Its definition coincides with the definition of the context split, except in the case of a linear storable type (that is, a type of the form $\ql\ \Pt$). In this case the occurrence of $\x:\ql\ \Pt$ in the $i$-th context is preceded by occurrences of $\x$ as a hidden object. In the following rule $j$ takes the values $1,...,n$.

\begin{center}
$
\begin{array}{c}
\Gamma_1\sqcup...\sqcup\Gamma_n=\Gamma\\
\overline{(\Gamma_1,\x: \qh\ \Pt) \sqcup... \sqcup(\Gamma_{j-1},\x: \qh\ \Pt)\sqcup(\Gamma_{j},\x: \ql\ \Pt)\sqcup\Gamma_{j+1}\sqcup...=\Gamma,\x:  \ql\ \Pt}\\
\end{array}
$
\end{center}

To express the fact that an argument of a basic operator can be an expression of type $\q\ \B_i$, or a variable of pseudotype $\qh\ \B_j$, we introduce the \textit{pseudotyping relation} $\ \Gamma\Vdash\e:\T$ as the extension of the relation $\Gamma\vdash\e:\T$ with the following rule: $\qu(\Gamma_1,\Gamma_2)$ implies $\Gamma_1,\x:\qh\ \B,\Gamma_2\Vdash\x:\qh \ \B$.

For handling patterns we will need the following notation. We use $[\p:\T]$ to denote the phrase of type $\Gamma$ which consists of flattening the pattern $\p$ and the type $\T$. That is, we define $\left[\x:\T\right] =  \x:\T$, if $\T=\q\ \Pt$ or $\T=\T_1\rightarrow\T_2$,  and

\medskip
$
\begin{array}{lll}
\left[\La\p_i,...,\p_n\Ra : \La\T_1,...,\T_n\Ra\right] & = & \left[\p_1 : \T_1\right],...,\left[\p_n : \T_n\right]\\
\end{array}
$

\medskip
The rules of the type system are given in Figures 5 and 6.

\begin{center}
\begin{tabular}{|c |}
\hline
$
\begin{array}{ll}
(\mathsf{var})
\begin{array}{l}
\quad \qu(\Gamma_1)\\
\quad \qu(\Gamma_2)\\
\overline{\Gamma_1, \x :\q\ \Pt,\Gamma_2\vdash \x :\q\ \Pt}\\
\end{array}
&
(\mathsf{bop})
\begin{array}{l}
\quad(\tau=(E_1,...,E_n)\rightarrow\E )\\
\quad\Gamma_i\Vdash \e_i:E_i\\
\overline{\Gamma_1\circ...\circ\Gamma_n\vdash \op^\tau(\e_1,...\ ):\E}\\
\end{array}
\medskip\\
(\mathsf{app})
\begin{array}{l}
\quad\Gamma\vdash \e:\T\\
\underline{\quad\Gamma\ \f=\T\rightarrow\T'}\\
\quad\Gamma\vdash \f\ \e:\T'\\
\end{array}
&
(\mathsf{tup})
\begin{array}{l}
\quad\Gamma_i\vdash \e_i:\T_i\\
\overline{\Gamma_1\sqcup...\sqcup\Gamma_n\vdash\qquad\qquad\ \ }\\
\qquad\La\e_1,...,\e_n\Ra:\La\T_1,...,\T_n\Ra\\
\end{array}
\medskip\\
(\mathsf{let})
\begin{array}{l}
\quad\Gamma_1\vdash \e:\T\\
\quad\left[ \p:\T\right] \vdash \p:\T\\
\quad\Gamma_2,\left[ \p:\T\right]\vdash \e':\T'\\
\overline{ \Gamma_1\sqcup\Gamma_2\vdash \Let\  \p\equiv\e\ \In\ \e':\T'}\\
\end{array}
&
(\mathsf{con})
\begin{array}{l}
\quad\Gamma_1\vdash \e:\q\ \Bool\\
\quad\Gamma_2\vdash \e_i:\T\\
 \overline{\Gamma_1\sqcup\Gamma_2\vdash\qquad\qquad\qquad\ \ }\\
 \qquad \If\ \e\ \Then\ \e_1\ \Else\ \e_2:\T\\
\end{array}
\medskip\\
(\mathsf{cas})
\begin{array}{l}
\quad\Gamma_1\vdash \e:\q\ [\E]\\
\quad\Gamma_2\vdash \e_0:\T\\
\quad\Gamma_2,\z_1:\E,\z_2:\q\ [\E]\vdash \e_1:\T\\
\overline{ \Gamma_1\sqcup\Gamma_2\vdash \Case^{\q}\  \e\ \Of\qquad\quad\ }\\
\qquad (\e_0,(\z_1:\z_2)\rightarrow\e_1)):\T\\
\end{array}
&
(\mathsf{bco})\ \
\begin{array}{l}
\quad\E_2=\q\ [\E_1]\\
\quad\sigma=(\E_1,\E_2)\rightarrow\E_2\\
\quad\q\ (\E_1)\\
\quad\Gamma_i\vdash\e_i:\E_i\\
\overline{\Gamma_1\sqcup\Gamma_2\vdash(\e_1:\e_2)^\sigma:\E_2}\\
\end{array}
\\
\end{array}
$\\
\hline
\end{tabular}\\
\tiny Figura 5: $\Gamma\vdash\e:\T$
\end{center}

Note that the predicates $\q(\T)$ and $\q(\Gamma)$ are completely removed from the typing rules, except for rules $(\mathsf{var})$ and $(:)$ (see  \cite{gramagliawlt}). This is because the tuples are not allocated in the store, and the functions are unrestricted.

\begin{center}
\begin{tabular}{|c |}
\hline
$
\begin{array}{l}
(\mathsf{sem})
\begin{array}{c}
\\
\underline{\quad\quad\quad\quad\quad}\\
\ \vdash []:[]\\
\end{array}
\quad\qquad\qquad\qquad\qquad\quad
(\mathsf{sba})
\begin{array}{c}
\\
\underline{\vdash\St: \Gamma\qquad(\op:\B)\in\Sigma}\\
\vdash \St,\x=\op:\Gamma,\x:\sq\  \B\\
\end{array}
\medskip\\
(\mathsf{sco})
\begin{array}{c}
\underline{\vdash\St: \Gamma_1\sqcup\Gamma_2 \qquad   \Gamma_1 \vdash (\x_1\!:\x_2)^\sigma: \E}\\
\vdash \St,\x=(\x_1\!:\x_2) : \Gamma_2,\x:\E\\
\end{array}
\medskip\\
(\mathsf{sfu})
\begin{array}{c}
\underline{\vdash\St: \Gamma \qquad  \left[\p:\T\right]\vdash \p:\T \qquad \Gamma^\qu,\f:\T\rightarrow\T', \left[\p:\T\right]\vdash \e:\T'}\\
\vdash \St,\f=\lambda \p.\e : \Gamma,\f:\T\rightarrow\T'\\
\\
\end{array}
\end{array}
$\\
\hline
\end{tabular}\\
\tiny Figura 6: $\vdash\St:\Gamma$
\end{center}

Note also that the definition of the relationship $\vdash\St:\Gamma$ no longer uses type context splitting, since the values $\op,\lambda\p.\e$ that are allocated in $\St$ cannot have linear free variables.

Finally, the relation $\vdash(\St,\e) $ is defined by the rule:

\medskip
\qquad\qquad\qquad\qquad$
\begin{array}{c}
\underline{\vdash\St: \Gamma\qquad \Gamma\vdash \e:\T}\\
\vdash (\St,\e)\\
\end{array}
$
\medskip

The condition $\vdash (\St,\e)$ (weak-linear typing) guarantees the correct linear evaluation (see \cite{gramagliawlt}), which we will defined in the next section.

Finally we note that the given inference rules are highly non-deterministic, due to the non-deterministic split (and pseudosplit) operation. Fortunately it is relatively easy to obtain a deterministic type checking  algorithm by using the free variables of the subphrases of an expression to determine the split operation (see \cite{walker}). We will use this idea for the linearization algorithm in section \ref{algoritmodelinealizacion}.

\subsection{Small-step semantic}
\label{semanticassmallstep}

Different ways of qualifying the list types $\sigma$ and the operator types $\tau$ will give rise to different forms of evaluation, which will differ in the way memory resources are managed.


To define small-step semantics we will use context-based semantics, whose distinctive characteristic is the explicit management of the store $\St$, for which we assume that no variables are repeated, and that when extending it, a new variable is used, supplied by $new\ \St$.

\textit{Evaluation context} $\e[]$ and the \textit{context rule} are defined in Figure 7.

\begin{center}
\begin{tabular}{|c |}
\hline
$
\begin{array}{l}
\begin{array}{lllllll}
(\St_0\e_0) \rightarrow_\beta (\St_1 \e_1)\\
\overline{(\St_0\e[\e_0]) \rightarrow (\St_1,\e[\e_1])}\\
\end{array}
\begin{array}{lllllll}
\e[] &::=  & []  \qquad\qquad\qquad\qquad\qquad\ \  \textsf{E-Contexts}\\
 &    & (\e[]:\e)^\sigma\\
 &    & (\x:\e[])^\sigma\\
 &    & \op^\tau(\x_1,...,\x_{i-1},\e[],...,\e_n)\\
&     &  \La \p_1,...,\p_{i-1},\e[],..,\e_n\Ra\\
     &     & \x\ \e[]\\
     &    &\If\ \e[]\ \Then\ \e_2\ \Else \ \e_3\\
     &     &\Let\ \p\equiv\e[]\ \In\ \e_2\\
     &     &\Case^\q\ \e[]\ \Of\ (\e_2,(\z_1\!\!:\!\z_2)\!\!\rightarrow\!\e_3)\\
\end{array}\\
\end{array}
$\\
\hline
\end{tabular}
\tiny Figura 7: Evaluation contexts and contexts rule.
\end{center}

By  $[\p\mapsto\p']$ we extend the substitution\footnote{$\x_1\mapsto\y_1$ denotes the identity map modified in the variable $ \x_1$, where it takes the value $\y_1$. } $\x\mapsto \y$ to patterns. Such extension is given by the conditions: $[\x\mapsto\y]= \x\mapsto \y$, $[\ep\mapsto \p]= []$ and

\medskip
$
\begin{array}{rlllll}
\left[\La \p_1,...,.\p_n\Ra\mapsto \La \p'_1,...,.\p'_n\Ra\right]  & = &
\left[\p_1\mapsto \p'_1\right],..., \left[\p_n\mapsto \p'_n\right]
\end{array}
$

\medskip

To represent memory deallocation we will use the operator $\sim_{\sq_1,...,\sq_n}$, defined by the following conditions:
\medskip

$
\begin{array}{rlll}
(\St,\x=\va,\St')\sim_\ql\x & = &\St,\St'\\
\St\sim_\sq\x & = &\St\qquad(\sq\neq\ql)\\
\St\sim_{[]}\mathsf{[]} & = &\St\\
\St\sim_{\sq,\sq\mathsf{s}}\x,\mathsf{xs} & = &(\St\sim_\sq\x)\sim_{\sq\mathsf{s}}\mathsf{xs}\\
\end{array}
$

\medskip

Terminal configurations will be pairs of the form $(\St,\p)$. We take the same program $(\St,\e)$ as the initial configuration. In Figure 8 the rules of the small-step semantics are given.

Note that the absence of qualifiers in the store limits the possibilities of destructive memory management, as shown by the rules ($\mathsf{eif})$\footnote{Boolean values are not destroyed, but they are not counted (see section \ref{resumendecasosdeestudio}).},$(\mathsf{eap})$\footnote{Functions are always unrestricted.} y $(\mathsf{ele})$\footnote{Tuples are not storable.}. We reiterate that the objective of these modifications with respect to the system presented in \cite{gramagliawlt} is to generate a theoretical framework for the safe introduction of global variables and updates in-place.

In the rules \textsf{eva} and \textsf{eop} the variable $\x$ is provided by the operator $new(\St)$.

\begin{center}
\begin{tabular}{|c |}
\hline
$
\begin{array}{lll}
(\mathsf{eva}) & (\St,\va) \rightarrow_\beta(\St,\x=\va,\x)& 
\smallskip\\
\mathsf{(eop)} & (\St,\op^\tau(\x_1,...,\x_n)) \rightarrow_\beta&(\St\x_i= \w_i, \\
&\quad (\St\sim_{\sq_1,...,\sq_n}\x_1,...,\x_n,\x=\va,\x)&\tau =( \sq_1\ \B_1,...\sq_n\ \B_n )\rightarrow\q\ \B)\\
& & \va=\op(\w_1,...,\w_n)\smallskip\\
(\mathsf{eif}) & (\St, \If\ \x\ \Then\ \e_0\ \Else\ \e_1)\rightarrow_\beta(\St,\e_0) & (\St\x=\True)\smallskip\\
 & (\St, \If\ \x\ \Then\ \e_0\ \Else\ \e_1)\rightarrow_\beta(\St,\e_1) & (\St\x=\False)\smallskip\\
(\mathsf{ele}) & (\St, \Let\ \p\equiv\ \p'\ \In\  \e) \rightarrow_\beta (\St, [\p\mapsto\p']\e) &\smallskip\\
(\mathsf{eap}) & (\St, \f\ \p') \rightarrow_\beta (\St, [\p\mapsto\p']\e) & (\St \f=\ (\lambda\p:\T.\e))\smallskip\\
\mathsf{(eem)} & (\St,[]) \rightarrow_\beta (\St,\x=[],\x)\smallskip\\
\mathsf{(eco)} & (\St,(\x_1:\x_2)^\sigma) \rightarrow_\beta(\St,\x=(\x_1:\x_2),\x)\smallskip\\
\mathsf{(eca)} & (\St,\Case^\q\ \x\ \Of\ (\e_0,(\z_1,\z_2)\rightarrow\e_1) \rightarrow_\beta & \\
 & \qquad(\St\sim_\q\x,\e_0)&\quad (\St\x=[])\\
 & \qquad(\St\sim_\q\x,[\langle\z_1,\z_2\rangle\mapsto\langle\x_1,\x_2\rangle]\e_j)& \quad(\St\x=(\x_1:\x_2))\smallskip\\
\end{array}
$\\
\hline
\end{tabular}
\tiny Figura 8: Small-step semantic
\end{center}

\section{Global programs}
\label{programasglobales}
The integration of the applicative and imperative paradigms into the same language presents important theoretical challenges. We will concentrate on the problem of obtaining imperative programs that use update in-place safely using the linear evaluation model.
For this, the next objective is to define a theoretical imperative language that adequately models in-place update. Our imperative language will admit different degrees of ``imperativeness'' for a program, which will be compared with linearity: we will look for linear programs that in their imperative version have the same cost in use of memory resources as the linear version.

The main property of the imperative language that we will now give is its theoretical simplicity. We obtain an imperative language $L^1[\Sigma^\g]$ by simply changing in $L^1[\Sigma^\q]$ the qualification of the basic operators.

Our imperative language is constructed from a qualified heterogeneous signature $\Sigma^\g$ formed by operators of the form $\op^\theta$, with $\theta=(\g_1\ \B_1,...,$ $\g_n\ \B_n)\rightarrow\g\ \B$. In Figure 8, $\Sigma^\g$ is defined.

\begin{center}
\begin{tabular}{|c |}
\hline
$
\begin{array}{llll}
\g & ::= & \lo\ |\ \x\  & \mathsf{global\ qualifiers}\smallskip\\
\B & ::= & \Int\ |\ \Bool\ |\  \Array\ |\ ...&  \mathsf{basic\ pretype}\smallskip\\
\Pt & ::= & \B\ |\ \left[\E\right]\ &  \mathsf{storable\ pretype}\smallskip\\
\E & ::= & \g\ \Pt&  \mathsf{storable\ type}\smallskip\\
\rho & ::= & \g\ [\E]\ |\ (\E,\g\ [\E])\rightarrow \g\ [\E]&  \mathsf{constructor\ type}\smallskip\\
\theta & ::= & (\g_1\ \B_1,...,\g_n\ \B_n)\rightarrow\g\ \B\ |&\mathsf{operators \ types}\smallskip\\
\Sigma^\g & ::= & \{(\op^\theta:\theta)\ :\ (\op:(\B_1,..., \B_n)\rightarrow\B)\in \Sigma\}\ \cup\ \  & \mathsf{qualified\ signature}\\
&  & \{([]^\rho:\rho)\}\ \cup\ \{(:)^\rho:\rho\}
\end{array}
$\\
\hline
\end{tabular}
\tiny Figura 8: Global Qualifiers and globally qualified signature\end{center}


The \textit{global} modality, given by the qualified type $\x\ \B$, will allow us to both add the in-place update and systematize the concept of a global variable.
An elementary way to do this is to consider the semantics given in the section \ref{semanticassmallstep}, and introduce phrases that, when evaluated, modify the memory $\St$ through update in-place $\St[\x]:=\va$.
The small-step semantics of $L^1[\Sigma^\g]$ is a slight modification of the one given for $L^1[\Sigma^\q]$. It only requires redefining the rules of the basic operators and constructors. In the following rules, let $\theta=(E_1,...,E_n)\rightarrow\g\ \B$, $\va=\op(\St\x_1,...,\x\cf_n)$ and $\rho=(\E_1,...,\E_n)\rightarrow\g\ \E$ (here $n\in\{0,2\}$).

\medskip
$
\begin{array}{lll}
(\mathsf{eol}) & (\St,\op^\theta(\x_1,...,\x_n)) \rightarrow (\St,\x=\va, \x)& (\g=\lo,\x=new(\St))\smallskip\\
(\mathsf{eog})& (\St,\op^\theta(\x_1,...,\x_n)) \rightarrow (\St[\x]:=\va, \x)& (\g=\x)\smallskip\\
(\mathsf{eel}) & (\St,[]^{\rho}) \rightarrow (\St,\x=[], \x)& (\g=\lo,\x=new(\St))\smallskip\\
(\mathsf{ecl}) & (\St,(\x_1:\x_2)^{\rho}) \rightarrow (\St,\x=(\x_1:\x_2), \x)& (\g=\lo,\x=new(\St))\smallskip\\
(\mathsf{eeg})& (\St,[]^\rho) \rightarrow (\St[\x]:=[], \x)& (\g=\x)\smallskip\\
(\mathsf{ecg}) & (\St,(\x_1:\x_2)^\rho) \rightarrow (\St[\x]:=(\x_1:\x_2), \x)& (\g=\x)
\end{array}
$

\medskip
This memory manipulation could be understood as a special case in the evaluation of $\op^\tau(\x_1,...,\x_n)$, with $\tau=(\sq_1\ \B_1,...,\sq_n\ \B_n)\rightarrow\ql\ \B$. If we have the situation $\sq_i\ \B_i=\ql\ \B$ and $\sq_j\neq\ql$ for $j\neq i$, then the evaluation deallocates $\x_i$, and claims a new memory address to alloc the result. If it were $\x_i$, then the effect is that of assignment $\x_i:=\op(\x_1,...,\x_n)$.
Since this relationship precisely defines the link between linearity and in-place updating, the concept of \textit{functional residual} will be relevant to determine how much of the efficiency of the linear version cannot be captured by the imperative version.

\subsection{Global type system}
\label{sistemadetipoglobal}
The type system of $L^1[\Sigma^\g]$ is intended to capture the imperative nature of the program. This will be reflected in the map  $P$ that we will define in section \ref{formaimperativa}.
It is a dependent type system that forces restricted use of variables of global type. Global types and global type  context are defined in Figure 9.

\begin{center}
\begin{tabular}{|c |}
\hline
$
\begin{array}{ll}
\begin{array}{llll}
\T & ::= & \E  \qquad\ \qquad \mathsf{Expression\  types}\\
&     & \Pi\p:\T_1.\T_2\\
&     & \La\T_1,...,\T_n\Ra
\\
\\
\end{array}
&
\begin{array}{rlllll}
\V & ::= & \E \qquad \qquad\mathsf{Value\  types}\\
&     & \Pi\p:\T_1.\T_2\smallskip\\
\Gamma    & ::=    & []\qquad\qquad \mathsf{Type\  context}\\
& & \Gamma,\ \x\ :\V\\
\end{array}
\end{array}
$\\
\hline
\end{tabular}
\tiny Figure 9: Global types and global type contexts
\end{center}

Note that we also use $\Gamma,\T$ to denote global contexts and global types. Depending on the signature involved ($\Sigma^\q$ or $\Sigma^\g$) we can distinguish whether we are talking about linear or global typing.

Before giving the rules of the global type system, let's give some technical definitions.
The $L^1[\Sigma^\g]$ types carry information about the store. The amount of information a type has is compared using the relation $\T\leq\T'$.
To define this relation, we define the poset $X$ as the lifting of the set of variables (with the flat order) with the smallest element $\lo$. The set of patterns $\p$ is embedded in the partially ordered set $Pat$ formed by the direct sum of all possible combinations of direct products of $X$\footnote{For example $\La \x,\La\z,\lo\Ra\Ra$ is an element of $\La X,\La X, X\Ra\Ra$, and $\La \x,\La\z,\lo\Ra\Ra\leq\La \x,\La\z,\w\Ra\Ra$ is verified (this last element is a supremum).}.

The relation $\T\leq\T'$ is determined by the condition $\T\leq\T' \Leftrightarrow\p_\T\leq\p_{\T'}$, where the map $\T\rightarrow\p_\T$ is defined by the following conditions:

\medskip

$
\begin{array}{rlll}
\p_{(\g\ \!\Pt)}& = &\g& \\
\p_{(\Pi\p:\T_1.\T_2)}& = &\lo\\
\p_{\La\T_1,...,\T_n\Ra}& = & \La\p_{\T_1},...,\p_{\T_n}\Ra\\
\end{array}
$

\medskip


In our type system, the function type $\T\rightarrow \T'$ is replaced by the type $\Pi\p\!:\!\!\T.\T'$, where the pattern $\p$ is a transmitter of information about the store that carries the argument of type $\T$. Then the type $\Pi\p\!:\!\!\T.\T'$ will only make sense when $\p$ can faithfully carry the information about the store that has the type $\T$. This condition is formalized by the property $\p_\T\leq\p$. This means that, for example, the type $\Pi \x:\z\ \Int.\T$ is meaningless to us.
 
In the application $\f\ \e$, with $\St\f=\lambda\p.\e_0$ and $\Gamma\f= \Pi\p\!:\!\!\T.\T'$, the pattern $\p$ (by a substitution\footnote{Trivially defined.}) will transmit to $\T'$ the information about the store carried by $\e$. This information to be transmitted will take the form of an object of $Pat$. We define $p^\Gamma\e\in Pat$ by the following conditions.
\medskip

\noindent$
\begin{array}{lcll}
p^\Gamma\ \x & = &  \x\\
p^\Gamma\ (\op^\theta(\e_1,...,\e_n))  & = & \g \qquad\qquad\qquad (\theta=(E_1,...,E_n)\rightarrow\g\ \Pt)\\
p^\Gamma\ []^\rho & = & \g \qquad\qquad\qquad (\rho=\g\ \Pt)\\
p^\Gamma\ (\e_1:\e_2)^\rho & = & \g \qquad\qquad\qquad (\rho=(E_1,\g\ \Pt)\rightarrow\g\ \Pt)\\
p^\Gamma\ (\If\ \e_1\ \Then\ \e_2\ \Else\ \e_3) & = &p^\Gamma\  \e_2\\
p^\Gamma\ (\Case\ \e_1\ \Of\ (\e_2,(\z_1:\z_2)\! \rightarrow\!\e_3)) & = &p^\Gamma\  \e_2\\
p^\Gamma\ \La\e_1,...,\e_n\Ra & = & \La p^\Gamma\ \e_1,...,p^\Gamma\ \e_n\Ra &\\
p^\Gamma\ (\Let\ \p\equiv\e_1\ \In\ \e_2) & = &[\p\mapsto p^\Gamma\e](p^\Gamma\  \e_2)\\
p^\Gamma\ (\f\ \e) & = &  \p_{[\p\mapsto p^\Gamma\e]\T'}\qquad\qquad\quad (\Gamma\f= \Pi\p\!:\!\!\T.\T')
\end{array}
$

\medskip

We now give the typing rules. The \textsf{loc} rule shown below presents the distinctive character of the type system, and expresses that the type takes from the variable (as a memory location) information about the store. The \textsf{var} rule is a typical typing rule for variables of basic types and lists.

\medskip

$\mathsf{(loc)}
\begin{array}{c}
\Gamma\vdash \x :\lo\ \Pt\\
\overline{\Gamma\vdash \x :\x\ \Pt}\\
\end{array}
\qquad\qquad\mathsf{(var)}
\begin{array}{c}
\\
\overline{\Gamma_{1},\x :\g\ \Pt,\Gamma_{2}\vdash \x :\g\ \Pt}\\
\end{array}
$

\medskip

\noindent
Note that \textsf{loc} allows an expression to be typed with several different types, all of them differing in the level of information about global variables that the type carries.
This feature is central to the overall system: a function, operation or constructor can receive data of the type $\T$ that corresponds to its specification, or data of type $\T'$ that carries more information than expected ($\T\leq\T'$). This is made possible by the rule:

\medskip

$\mathsf{(\leq)}
\begin{array}{c}
\underline{\Gamma\vdash \e :\T'\quad \T\leq\T'}\\
\Gamma\vdash \e :\T\\
\end{array}
$

\medskip

The rest of the rules are given in Figure 11 and 12. Below we define the context operator $\Gamma_1;\Gamma_2$, used in Figure 11.

\begin{center}
\begin{tabular}{|c |}
\hline
$
\begin{array}{ll}
(\mathsf{bop})
\begin{array}{l}
\quad\theta=(\E_1,...,\E_n)\rightarrow\E\\
\quad\Gamma\vdash \e_i:\E_i\\
\overline{\Gamma\vdash \op^\theta(\e_1,...,\e_n ):\E\quad}\\
\end{array}
&
(\mathsf{bco})\ \
\begin{array}{l}
\quad\rho=(\E_1,\E_2)\rightarrow \E_2\\
\quad\Gamma\vdash\e_i:\E_i\\
\overline{\Gamma\vdash(\e_0:\e_1)^\rho:\E_2\quad}\\
\end{array}
\medskip\\
(\mathsf{tup})
\begin{array}{l}
\quad\Gamma\vdash \e_i:\T_i\\
\overline{\Gamma\vdash\La\e_1,...,\e_n\Ra:\La\T_1,...,\T_n\Ra}
\end{array}
&
(\mathsf{em})
\begin{array}{l}
\quad\rho=\E\\
\overline{ \Gamma\vdash []^\rho:\E}\\
\end{array}
\medskip\\
(\mathsf{cas})
\begin{array}{l}
\quad\Gamma\vdash\e:\lo\ [\lo\ \Pt]\\
\quad\Gamma\vdash \e_1:\T\\
\quad\Gamma,\ \! \z_1\!\!:\!\lo\ \Pt,\ \! \z_2\!\!:\!\lo\ \! [\lo\ \Pt]\vdash \e_2:\T\\
\overline{ \Gamma\vdash \Case^{\lo}\  \e\ \Of\
(\e_1,(\z_1\!\!:\!\z_2)\!\!\rightarrow\e_2):\T}\\
\end{array}
&
(\mathsf{app})
\begin{array}{l}
\quad\Gamma\ \f=\Pi\p:\T.\T'\\
\quad\Gamma\vdash \e: \T \\
\overline{\Gamma\vdash \f\ \e:[\p\mapsto p^\Gamma\e]\T'}\\
\end{array}
\medskip\\
(\mathsf{let})
\begin{array}{l}
\quad\Gamma\vdash \e:\T\\
\quad[\p\!:\!\!\T]\vdash\p: \T\\
\quad\Gamma; [\p\!:\!\! \T]\vdash \e':\T'\\
\overline{\Gamma\vdash \Let\  \p\equiv\e\ \In\ \e':[\p\mapsto p^\Gamma\e]\T'}\\
\end{array}
&
(\mathsf{con})
\begin{array}{l}
\quad\Gamma\vdash \e:\g\ \Bool\\
\quad\Gamma\vdash \e_i:\T\\
 \overline{\Gamma\vdash \If\ \e\ \Then\ \e_1\ \ }\\
\qquad\quad\quad  \Else\ \e_2:\T\\
\end{array}
\\
\end{array}
$\\
\hline
\end{tabular}\\
\tiny Figura 11: $\Gamma\vdash\e:\T$
\end{center}

Well-typed  $L^1[\Sigma^\g]$-programs will be called \textit{global}.






The existence of global variables in the context $\Gamma$ forces a restricted handling of bound variables in $\Let$ statement. Classical type systems use the environment operation $\Gamma_1,\Gamma_2$, where a variable that occurs in both contexts is redefined by $\Gamma_2$ overriding its original definition set in $\Gamma_1$.
In the global system this operation must be restricted, becoming a partially defined operation: a global variable (i.e. of type $\x\ \Pt$) that occurs in $\Gamma_1$ cannot be redefined.
In the following definition we use the predicates $\lo(\V)$, which is defined by the following conditions: $\lo(\T\rightarrow\T')=\True$,  $\lo(\lo\ \Pt)=\True$,  $\lo(\x\ \Pt)=\False$  and $\gl(\V)=\neg\lo(\V)$.

\medskip

$
\begin{array}{rlll}
\Gamma_1;[] & = & \Gamma_1\\
(\Gamma^1_1,\x:\V_1,\Gamma_1^2);(\x:\V_2,\Gamma_2) & = & (\Gamma^1_1,\x:\V_1,\Gamma_1^2);\Gamma_2 & \textsf{if }\gl(\V_1)\wedge\gl(\V_2)\\
(\Gamma^1_1,\x:\V_1,\Gamma_1^2);(\x:\V_2,\Gamma_2) & = & (\Gamma^1_1,\Gamma_1^2,\x:\V_2);\Gamma_2 & \textsf{if }\lo(\V_1)\\
(\Gamma^1_1,\x:\V_1,\Gamma_1^2);(\x:\V_2,\Gamma_2) & = & (\Gamma^1_1,\Gamma_1^2);\Gamma_2 & \textsf{if }\gl(\V_1)\wedge\lo(\V_2)\\
\end{array}
$

\medskip

The rule for  $\Let$-construction reports the existence of a global variable. Indeed, if $\Gamma \x=\x\ \Pt$ and $\x$ occurs in $\p$, then the restriction imposed by the operator $(;)$ forces the type $\x\ \Pt$ for the data corresponding to $\x$.
In the case of the rule \textsf{cas} we do not have this possibility, since the parameters are forced to be as uninformative as possible.
The need for this restriction can be seen in the example \textsf{case} (see at the end of the section \ref{glosariodecasosdeestudio}).

\begin{center}
\begin{tabular}{|c |}
\hline
$
\begin{array}{l}
(\mathsf{sem})
\begin{array}{c}
\\
\underline{\quad\quad\quad\quad\quad}\\
\ \vdash [ ]:[]\\
\end{array}
\quad\qquad\qquad\qquad
(\mathsf{sba})
\begin{array}{c}
\\
\underline{\vdash\St : \Gamma\qquad \Gamma\vdash \op^{\g\ \!\B}:\g\ \!\B}\\
\vdash \St,\x=\op:\Gamma,\x:\g\ \!\B\\
\end{array}
\medskip\\
(\mathsf{sco})
\begin{array}{c}
\underline{\vdash\St : \Gamma \qquad   \Gamma \vdash (\x_1\!\!:\!\x_2)^\rho: \E}\\
\vdash \St,\x=(\x_1\!\!:\!\x_2) : \Gamma,\x:\E\\
\end{array}
\medskip\\
(\mathsf{sfu})
\begin{array}{c}
\underline{\vdash\St: \Gamma \qquad  \left[\p:\T\right]\vdash \p:\T \qquad \Gamma,\f:\Pi\p:\T.\T', \left[\p:\T\right]\vdash \e:\T'}\\
\vdash \St,\f=\lambda \p.\e: \Gamma,\f:\Pi\p:\T.\T'\\
\\
\end{array}
\end{array}
$\\
\hline
\end{tabular}\\
\tiny Figura 12: $\vdash\St:\Gamma$
\end{center}

\medskip

We end the section by pointing out that typing is strongly nondeterministic, due to the rule $(\leq)$. One possible implementation is to restrict its application to arguments of function calls (rule \textsf{app}), constructors (\textsf{bco}) and basic operators (\textsf{bop}). These modifications are shown below.

\medskip

\noindent\!\!$\begin{array}{l}
\quad\Gamma\ \f=\Pi\p:\T_0.\T'\\
\quad\Gamma\vdash \e: \T \\
\quad\T_0\leq\T\\
\overline{\Gamma\vdash \f\ \e:[\p\mapsto p^\Gamma\e]\T'}\\
\end{array}$
\quad
$
\begin{array}{l}
\quad \g_i=p^\Gamma \e_i\\
\quad\rho=(\g^0_1\ \Pt^0_1,\g^0_2\ \Pt^0_2)\!\rightarrow\!\E\\
\quad\Gamma\vdash\e_i:\g_i\ \Pt_i\\
\overline{\Gamma\vdash(\e_0:\e_1)^\rho:\E\qquad\quad}\\
\end{array}
$
\quad
$\begin{array}{l}
\quad \g_i=p^\Gamma \e_i\\
\quad\theta=(...,\g^0_i\ \B_i,...)\!\rightarrow\!\E\\
\quad\Gamma\vdash \e_i:\g_i\ \B_i\\
\overline{\Gamma\vdash \op^\theta(\e_1,...,\e_n ):\E\quad}\\
\end{array}
$
\medskip

\noindent
\noindent In the last two rules, note that $\g_i^0\leq\g_i$ (in $Pat$), then we have $\Gamma\vdash\e_i:\g^0_i\ \Pt_i$.

\subsection{Imperative form for global programs}
\label{formaimperativa}

In this section we will show that a well-typed $L^1[\Sigma^\g]$ program can take an imperative form, which consists of making assignment and global variables explicit.

The operator $P\ \Gamma\ \e$ will be defined for expressions $L^1[\Sigma^\g]$ that satisfy $\Gamma\vdash\e:\T$ for some $\T$. We call $Gl\ \Gamma$ the set of variables that occur in $\bigcup_{\x\in dom\Gamma}FV\ \!(\Gamma\x)$. En la siguente definición asumimos que tanto $\rho$ como $\theta$ tienen la forma $(\g_1\ \Pt_1,...,\g_n\ \Pt_n)\rightarrow\g\ \Pt$ (para $\rho$ tenemos dos posibilidades: $n=0$ o $n=2$).

\medskip
\noindent$
\begin{array}{rclll}
P\ \Gamma\ \x & = & \x\smallskip \\
P\  \Gamma\ \op^\theta(\e_1,...,\e_n) & = & \op(P\  \Gamma\ \e_1,...,\ P\  \Gamma\ \e_n) & (\g=\lo)\smallskip\\
P\  \Gamma\ \op^\theta(\e_1,...,\e_n) & = & \x:=\op(P\  \Gamma\ \e_1,..., P\  \Gamma\ \e_n)& (\g=\x)\smallskip\\
P\  \Gamma\ []^\rho & = & [] & (\g=\lo)\smallskip\\
P\  \Gamma\ (\e_1:\e_2)^\rho & = & (P\  \Gamma\ \e_1:\ P\  \Gamma\ \e_2) & (\g=\lo)\smallskip\\
P\  \Gamma\ []^\rho & = & \x:=[] & (\g=\x)\smallskip\\
P\  \Gamma\ (\e_1:\e_2)^\rho & = & \x:=(P\  \Gamma\ \e_1:\ P\  \Gamma\ \e_2)& (\g=\x)\\
\end{array}
\\
$

\medskip

\noindent 
For all other phrases, $P$ only works by preventing global variables from being passed as parameters. In the first equation, $\pp$ denotes the pattern $\p$ in which the variables of $Gl\ \Gamma$ are replaced by $\ep$, leaving the rest unchanged.

\medskip

\noindent$
\begin{array}{rllll}
P\  \Gamma\ (\Let\ \p \equiv \e\ \In\ \e') & = & \Let\ \pp\equiv P\  \Gamma\ \e\ \In\ P\  \Gamma\ \e'   \smallskip\\
P\ \Gamma\ (\f\ \e ) & = & \f\ (P\   \Gamma\ \e)  \smallskip\\ 
P\  \Gamma\ \La\e_1,...,\e_n\Ra & = & \La P\  \Gamma\ \e_1,...,P\   \Gamma\ \e_n\Ra \smallskip \\
P\  \Gamma\ (\If\ \e\ \Then\  \e'\ \Else\ \e'' ) & = &\If\ P\  \Gamma\ \e\ \Then\  P\  \Gamma\ \e'\ \Else\ P  \Gamma\ \e'' \smallskip \\
P\  \Gamma\ (\Case\ \e\ \Of\  (\e_0,(\z\!:\!\zs)\rightarrow\e_1)) & = &\Case\ P\  \Gamma\ \e\ \Of\  (P\  \Gamma\ \e_0, (\z\!:\!\zs)\rightarrow\!P\  \Gamma\ \e_1)  \\
\end{array}
$

\medskip

\noindent 
Finally, for valued we define
$P\ \Gamma\ \va = \va$, if $\va$ is not a function type value, and $P\ \Gamma\ (\lambda \p.\e) = (\lambda \pp.\ P\ \Gamma\ \e)$.

\section{Linearization algorithm}
\label{algoritmodelinealizacion}

We will use the term \textit{trivial} to refer to phrases of type $\Gamma$, $\tau$, $\sigma$ or $\Sigma^\q$ whose qualifiers are all $\qu$. It should be noted that under this assumption, the system given in the section \ref{unsistemaslinealparal1sigma} constitutes a classical type system.

We give in this section an algorithm that takes a context of types $\Gamma$ and an expression $\e$ of $L^1[\Sigma^\qu]$, and returns a set $\mathbf{S}\ \Gamma \ \e$ of pairs of the form $(\e_i,\bt_i)$. These pairs represent all possible non-trivial qualifications of the operators of $\e$, followed by the resulting type of the expression.
Each pair $(\e_i,\bt_i)$ satisfies $\Gamma\vdash\e_i:\bt_i$. Specifically, what the algorithm does is replace each $\op^\tau$ (with trivial $\tau$), by $\op^{\tau'}$, where $\tau'$ represents the possible ways to put qualifiers to $\tau$. The same for each occurrence of $[]^\sigma$ and $(:)^\sigma$.

The operator $pspl\ n\ (X_1,...,X_n)\ \Gamma$ (context pseudosplit) is defined by the following rules. Here $X_j$ represent sets of variables.

\smallskip

$pspl\ n\ (X_1,...,X_n)\ [] \ =\ ([],...,[])$
\smallskip

\noindent 
Suppose that $pspl\ n\ (X_1,...,X_n)\ \Gamma=(\Gamma_1,...,\Gamma_n)$. If $\V=\q\ \Pt$, with $\q\neq\ql$, or $\V= \T_0\rightarrow\T_1$,  then:

\smallskip

$pspl\ n\ (X_1,...,X_n)\ (\Gamma,x:\V) \ = \ ((\Gamma_1,x:\V),...,(\Gamma_n,x:\V))$

\smallskip

\noindent 
To complete the definition of $pspl$ we add $n$ equations, which are obtained from the generic equation given below, when $i$ takes the values $1,...,n$. Here $i$ represents the largest $k$ such that $\x\in X_k$.

\smallskip

$
\begin{array}{rlll}
pspl\ n\ (X_1,...,X_n)\ (\Gamma,\x:\ql\ \B) & = & ((\Gamma_1,\x:\qh\ \B),...,(\Gamma_{i-1},\x:\qh\ \B),\smallskip\\
                                       &   & \qquad\qquad (\Gamma_{i},\x:\ql\ \B),\Gamma_{i+1},...,\Gamma_{n})\smallskip\\
\end{array}
$
\smallskip

\noindent 
If $\x\notin X_k$ for all $k$, then we define
\smallskip

$pspl\ n\ (X_1,...,X_n)\ (\Gamma,\x:\ql\ \B)=((\Gamma_1,\x:\ql\ \B),\Gamma_2,...,\Gamma_{n})$.

\smallskip

\noindent In a similar way, the operator $spl\ n\ (X_1,...,X_n)\ \Gamma$ can be defined, which differs from $pspl$ in that $\qh\ \B$ is not added in the contexts of the left when entering $\ql\ \B$.

We now give the definition $\mathbf{S}\ \Gamma\ \e$. Let us consider the set of qualifiers $\{\ql,\qu\}$ as a partially ordered set, with the relation $\leq$ defined by $\ql\leq\qu$, $\ql\leq\ql$ and $\qu\leq\qu$.

The variable expression case is given by: $\mathbf{S}\ \Gamma\ \x \ =\ [(\x,\Gamma\x)]$ if $\x\in dom(\Gamma)$. Otherwise we define $\mathbf{S}\ \Gamma\ \x \ =\  []$. To define the case $\mathbf{S}\ \Gamma\ \op^\tau(\e_1,...,\e_n)$, we introduce the following notation. Suppose that

\smallskip
$spl\ n\ (FV\ \e_1,...,FV \e_n)\ \Gamma \ =\ (\Gamma_1,...,\Gamma_n)$
\smallskip

\noindent and that

\smallskip

$\mathbf{S}\ \Gamma_i\ \e_i \ = \ [(\e^1_i,\E^1_i),...,(\e^{k_i}_i,\E^{k_i}_i)]$,

\smallskip

\noindent for $i=1,...,n$.

Given $\op^\tau$, with $\tau=(\E^0_1,...,\E^0_n)\rightarrow\q^0\ \B$ (it is not necessary to assume that the initial signature is trivial), for each tuple $J=(j_1,...,j_n)$, with $J\in[1,k_1]\times ...\times [1,k_n]$, we define, for $i=1 ,...,n$,

\smallskip

$
\begin{array}{llll}
\E^J_i  & = & \E^{j_i}_i\\
\e^J_i & = &  \e^{j_i}_i\\
\tau^{J,\q} & = & (\E^J_1,...,\E^J_n)\rightarrow\q\ \B
\end{array}
$

\smallskip

\noindent 
Then, the equation for phrases of the form $\op^\tau(\e_1,...,\e_n)$ is as follows:

\smallskip

\noindent $
\begin{array}{lllll}
\mathbf{S}\ \Gamma\ \op^\tau(\e_1,...,\e_n) & = & [\ (\op^{\tau^{J,\q}}(\e^J_1,...,\e^J_k),\q\ \B):& \q\leq\q^0,\smallskip\\
                                       &   &                                       & J\in[1,k_1]\times ...\times [1,k_n]\ ]\smallskip\\
\end{array}
$

\smallskip

The linearization algorithm treats constructors as operators, except that we use $pspl$ instead of $spl$.
Let $\sigma=(\E,\q_0\ [\E])\rightarrow \q_0\ [\E]$.
To define the case $\mathbf{S}\ \Gamma\ (\e_1:\e_2)^\sigma$, suppose that
$pspl\ n\ (FV\ \e_1,FV \e_2)\ \Gamma \ =\ (\Gamma_1,\Gamma_2)$ and that

\smallskip

$\mathbf{S}\ \Gamma_i\ \e_i \ = \ [(\e^1_i,\E^1_i),...,(\e^{k_i}_i,\E^{k_i}_i)]$,

\smallskip

\noindent for $i=1,2$. Para cada  $J=(j_1,j_2)$, con $J\in[1,k_1]\times[1,k_2]$, definimos

\smallskip

$
\begin{array}{llll}
\E^J_i  & = & \E^{j_i}_i\\
\e^J_i & = &  \e^{j_i}_i\\
\sigma^{J} & = & (\E^J_1,\E^J_2)\rightarrow\E^J_2
\end{array}
$

\smallskip

\noindent 
Then, the equation for phrases of the form $(\e_1:\e_2)^\sigma$ is as follows:

\smallskip

\noindent $
\begin{array}{lllll}
\mathbf{S}\ \Gamma\ (\e_1:\e_2)^\sigma & = & [\ ((\e^J_1:\e^J_2)^{\sigma^{J}},\E^J_2):& J\in[1,k_1]\times [1,k_2]\ ]\smallskip\\
\end{array}
$

\smallskip
Equations for conditional, tuple, and application can be obtained in a trivial way, constructing the results using all possible combinations of the results of the immediate subphrases. We complete the definition of the operator $\mathbf{S}\ \Gamma\ \e$ for phrases \textsf{let} and \textsf{case} with the following equations.

\smallskip

\noindent $\mathbf{S}\ \Gamma\ (\Let\ \p\equiv \e_1\ \In\ \e_2)  =  [\ (\Let\ \p\equiv \e^j_1\ \In\ \e^{j,i}_2,\T^{j,i}_2):  j\in[1,k_1],\ i\in [1,k_j]\ ]$

\smallskip

\quad 
\quad where  $pspl\ 2\ (FV\ \e_1,FV \e_2-FV\ \p)\ \Gamma \ =\ (\Gamma_1,\Gamma_2)$ and

\smallskip
$
\quad\begin{array}{rlll}
\mathbf{S}\ \Gamma_1\ \e_1 & = & [(\e^1_1,\T^1_1),...,(\e^{k_1}_1,\T^{k_1}_1)] \\
\mathbf{S}\ (\Gamma_2,\p:\T^j_1)\ \e_2& = & [(\e^{j,1}_2,\T^{j,1}_2),...,(\e^{j,k_j}_2,\T^{j,k_j}_2)] \qquad (j\in[1,k_1])\\
\end{array}
$

\smallskip

\noindent $\mathbf{S}\ \Gamma\ (\Case\ \e_1\ \Of\ ( \e_2,(\z_1\!:\!\z_2)\!\mapsto\!\e_3)  =  $\smallskip

\qquad$[\ (\Case\ \e^j_1\ \Of\ ( \e^r_2,(\z_1\!:\!\z_2)\!\mapsto\!\e^{j,i}_2,\T^{j,i}_2):  j\in[1,k_1],\ r\in[1,k_2],\ i\in [1,k_j]\ ]$

\smallskip

\quad where  $pspl\ 2\ (FV\ \e_1,FV\ \e_2\cup (FV \e_3-\{\z_1,\z_2\}))\ \Gamma \ =\ (\Gamma_1,\Gamma_2)$ and

\smallskip
\quad$
\begin{array}{rlll}
\mathbf{S}\ \Gamma_1\ \e_1 & = & [(\e^1_1,\T^1_1),...,(\e^{k_1}_1,\T^{k_1}_1)] \\
\mathbf{S}\ \Gamma_2\ \e_2 & = & [(\e^1_2,\T^1_2),...,(\e^{k_2}_2,\T^{k_2}_2)] \\
\mathbf{S}\ (\Gamma_2,\p:\T^j_1)\ \e_2& = & [(\e^{j,1}_2,\T^{j,1}_2),...,(\e^{j,k_j}_2,\T^{j,k_j}_2)] \qquad (j\in[1,k_1])\\
\end{array}
$

\smallskip

A simple way to obtain an improvement in the efficiency of this algorithm\footnote{Clearly the algorithm is exponential. The attempt to linearize  insertion sort with the algorithm without any improvement exceeds $2^{13}$ linearizations.} is to start from a non-trivial expression that sets certain qualifiers as linear, thus reducing the complexity of the search. The problem with this improvement is that if there is too much data that cannot be linearized, then the algorithm still has a high complexity. In this case, it is convenient to use the qualifier $\overline{\qu}$, with the meaning of a fixed qualifier, which the algorithm will not modify. Of course, these options involve setting conditions that restrict the typing possibilities. If conditions are set that result in a type inconsistency, the algorithm will return an empty list.

\section{Globalization Algorithm}
\label{algoritmodeglobalizacion}

In this section $\Gamma$, $\T$ represent respectively environment and qualified type according to the system given in \ref{programasglobales}. In the same way as when we study substructurality,
we use the term \textit{trivial} to refer to a signature $\Sigma^\g$, a type context $\Gamma$,
or a type $\T$, whose qualifiers are all $\lo$. It should be noted that under this hypothesis,
the system given in section \ref{programasglobales} constitutes classical type systems.

We give now an algorithm that takes a context of types $\Gamma$, an expression $\e$ of $L^1[\Sigma^\lo]$ and a global type $\T$, and returns a set $\mathbf{G}\ \Gamma \ \e\ \T$ of expressions $\e_i$. These represent all possible non-trivial qualifications of the operators of $\e$.
Each expression $\e_i$ satisfies $\Gamma\vdash\e_i:\T$. Specifically, what the algorithm does is replace each $\op^\theta$ (with trivial $\theta$), by $\op^{\theta'}$, where $\theta'$ represents the possible ways to put qualifiers to $\theta$. The same for each occurrence of $[]^\rho$ and $(:)^\rho$.

Unlike the linearization algorithm, it is essential here to pass a on-trivial type $\T$, in addition to the non-trivial type environment. The type $\T$ will allow us to express information that will condition the possible ways of qualifying operators and constructors.

In the definition of \textbf{G} we will use the function $\mathcal{T}_\Gamma \e$, which returns a type $\T$ satisfying $\Gamma\vdash \e:\T$. It is derived from the type checking algorithm given in section \ref{sistemadetipoglobal}.
We will also use the following notation. By $Gl^\Pt\ \Gamma$ we will denote the set of variables $\x$ that satisfy $\Gamma\x=\x\ \Pt$. We extend the notation as follows: if $\E=\g\ \Pt$, then by $Gl^\E\ \Gamma=Gl^\Pt\ \Gamma$ if $\g=\lo$, and $Gl^\E\ \Gamma=\{\x\}\cap Gl^\Pt\ \Gamma$, if $\g=\x$.

The operator $\mathbf{G}\ \Gamma\ \e\ \T$ is defined by the following conditions:

\medskip

$
\begin{array}{llll}
\mathbf{G}\ \Gamma\ \x \ \E& = & [\x] \ \ &(\Gamma\x\leq\E)\smallskip\\
\mathbf{G}\ \Gamma\ \x \ \E& = & []             \ \ & cc\smallskip\\
\end{array}
$

\medskip

\noindent 
To define $\mathbf{G}$ in the basic operators, we introduce the following notation.
Suppse  $\theta=(\lo\ \B_1,...,\lo\ \B_k)\rightarrow\lo\ \B$ and $\mathcal{T}_\Gamma\e_i=\g_i\ \B_i$. Define $C_i = \{\g_i\}\cup Gl^{\g_{i} \B_i}\ \Gamma$. Note that if $\g_i=\x$, then $C_i = \{\x\}$. Given $\g$,  for each $\gamma\in C_1\times...\times C_n$, we define:

\medskip

$
\begin{array}{llll}
\theta^{\gamma,\g}  & = & (\gamma_1\ \B_1,...,\gamma_n\ \B_n)\rightarrow\g\ \B  \\
\end{array}
$

\medskip

\noindent 
Then the equation for phrases of type $\op^\theta(\e_1,...,\e_n)$ is as follows:

\medskip

\noindent $
\begin{array}{lllll}
\mathbf{G}\ \Gamma\ \op^\theta(\e_1,...,\e_n) \ (\g\ \B)& = & [\ \op^{\theta^{\gamma,\g}}(\e'_1,...,\e'_k) \ | &
      \e'_j\in \mathbf{G}\ \Gamma\ \e_j\ (\gamma_j\ \B_j),\ \gamma]\smallskip\\
\end{array}
$

\medskip

\noindent Suppose now that $\rho=(\lo\ \Pt,\lo\ [\lo\ \Pt])\rightarrow \lo\ [\lo\ \Pt]$. Given $\g$ we define $\rho^{\g}   = (\lo\ \Pt,\lo\ [\lo\ \Pt])\rightarrow  \g\ [\lo\ \Pt]$.

\medskip

\noindent 
Then the equation for the list constructors $[]^{\lo\ \Pt}$ and $(\e_1:\e_2)^\rho$ are as follows:

\medskip

\noindent $
\begin{array}{lllll}
\mathbf{G}\ \Gamma\ []^{\lo\ \!\Pt} \ (\g\ \Pt)& = & []^{\g\ \!\Pt}\\
\mathbf{G}\ \Gamma\ (\e_1:\e_2)^{\rho_:} \ (\g\ \E)& = & [\ (\e'_1:\e'_2)^{\rho^{\g}} \ | &
      \e'_1\in \mathbf{G}\ \Gamma\ \e_1\ (\lo\ \Pt),\\
      		& & & \e'_2\in \mathbf{G}\ \Gamma\ \e_2\ (\lo\  [\lo\ \Pt]),\ \gamma]\\
\end{array}
$

\medskip

\noindent 
For tuples and the conditional we have:

\medskip

\noindent $
\begin{array}{lllll}
\mathbf{G}\ \Gamma\ \La\e_1,...,\e_k\Ra \La\T_1,...,\T_k\Ra& = & [\ \La\e'_1,...,\e'_n\Ra\ | \
                                               e'_j\in\mathbf{G}\ \Gamma\ \e_j\  \T_j\ ]\smallskip\\
\mathbf{G}\ \Gamma\ (\If\ \e\ \Then\ \e_1\ \Else\ \e_2)\ \T & = & [\ \If\ \e'\ \Then\ \e'_1\ \Else\ \e'_2\ |\ \smallskip\\
                                                       &   & \ \quad \e'_i\in\mathbf{G}\ \Gamma\ \e_i\  \T,\                              \e'\in\mathbf{G}\ \Gamma\ \e\  (\lo\ \Bool)                                                                     ]\\
\end{array}
$

\medskip

Globalizing an expression within the scope of a binding requires globalizing the environment to maximize the chances of success.
To implement this idea, the operator $\p\cdot\!\T$ will be useful, (partially) defined for $\p\in Pat$ by the following rules:

\smallskip

\qquad $
\begin{array}{rllll}
\x\cdot(\g\ \Pt) & = & \x\ \Pt\smallskip\\
\lo\cdot\T & = & \T\smallskip\\
\La\p_1,...,\p_n\Ra\cdot\La\T_1,...,\T_n\Ra & = &\La\p_1\cdot\T_1,...,\p_n\cdot\T_n\Ra\\
\end{array}
$

\medskip

\noindent 
We also introduce the following notation. If $A$ is a set of variables, we denote by $\p|_A$ the element of $Pat$ that is obtained by replacing the variables of $\p$ that are not in $A$ by $\lo$.
In the phrases application and $\Let$ we will use the notation $\pp=\p|_{Gl\ \Gamma}$. For the application we define:

\medskip

\noindent $
\begin{array}{lllll}
\mathbf{G}\ \Gamma\ (\f\ \e) & = & [\ \f\ \e' \ |\ \e'\in\mathbf{G}\ \Gamma\ \e\  (\pp\!\cdot\!\!\T),\ \Gamma\f=\Pi\p\!:\!\!\T.\T_1 ]\\
\end{array}
$

\medskip

\noindent 
Let $\T=\mathcal{T}_\Gamma\ \e$. Then:

\medskip

\noindent $
\begin{array}{lllll}
\mathbf{G}\ \Gamma\ (\Let\ \p\equiv \e\ \In\ \e_1)\ \T_1 & = & [\ \Let\ \p\equiv \e'\ \In\ \e'_1\ | &
 \!\!\!\!\e'\in\mathbf{G}\  \Gamma\ \e\ (\pp\!\cdot\!\T),\smallskip\\                                                         & & & \!\!\!\!\e'_1\in\mathbf{G}\ (\Gamma,[\p:\pp\!\cdot\!\T])\ \e_1\  \T_1   ]
\end{array}
$

\medskip
\noindent
Finally, assuming $\mathcal{T}_\Gamma\ \e_0=\q\ [\T_0]$, we have the equation for $\Case$:

\bigskip

\noindent $
\begin{array}{lllll}
\mathbf{G}\ \Gamma\ (\Case\ \q\ \e_0\ \Of\ (\e_1,(\z:\zs)\rightarrow\e_2)\ \T \!\!& \!\! = \ [& \!\!\!\!\Case\ \lo\  \e'_0\ \Of\ (\e'_1,(\z:\zs)\rightarrow\e'_2)\ | \smallskip\\
& & \!\!\e'_0\in\mathbf{G}\  \Gamma\ \e_0\ \T_0,\                                                        \e'_1\in\mathbf{G}\ \Gamma\ \e_1\  \T   \smallskip\\
& & \!\!\e'_2\in\mathbf{G}\ (\Gamma,\z\!:\!\T_0,\zs\!:\!\lo\ \![\T_0])\ \e_2\  \T  ]\\
\end{array}
$

\medskip

In the store, globalization only modifies the functions, but this modification is key for the system to work. Roughly speaking, in the body of a function, the variables that occur freely in the image must be globalized.
We will call $\Gamma^X$ the result of globalizing in $\Gamma$ the variables that occur in $X$. More precisely:

\medskip

\noindent $
\begin{array}{lllll}
\Gamma^X \x & = \Gamma\ \x& \textsf{if}\ \x\in dom\ \!\Gamma \ \textsf{and}\ \x\notin X\smallskip\\
\Gamma^X \x & = \x\ \Pt& \textsf{if}\ \x\in dom\ \!\Gamma,\ \x\in X\ \textsf{and}\ \Gamma\x=\q\ \Pt \smallskip\\
\Gamma^X \f & = \Pi\p\!:\!\pp\!\cdot\!\!\T_d.\T_i& \textsf{if}\ \x\in dom\ \!\Gamma\ \textsf{and}\  \Gamma\f=\Pi\p\!:\!\!\T_d.\T_i, \ \textsf{where}\ \pp=\p|_{X\cap FV\ \!\!\T_i}\\
\end{array}
$

\medskip
\noindent 
The globalization $\mathbf{G}\ \Gamma\ \St$ of the store $\St$ consists of replacing each function definition $\f=\lambda\p.\e$, by

\smallskip
$\f=\lambda\p.\mathbf{G}\ \Gamma_1^{FV\ \!\!\T_i}\ \e\ \T_i$,

\smallskip

\noindent where   $\Gamma\f=\Pi\p\!:\!\!\T_d.\T_i$ and $\Gamma_1=\Gamma,[\p\!:\!\!\T_d]$.


\section{Case study summary}
\label{resumendecasosdeestudio}

Concentrating memory management on the signature is the key to being able to compare the three forms of evaluation:  unrestricted, linear and global (imperative).

Given $\alpha=(\q_1\ \Pt_1,...,\q_n\ \Pt_n) \rightarrow\q\ \Pt$ and $\beta=(\g_1\ \Pt_1,...,\g_n\ \Pt_n)\rightarrow\g\ \Pt$, we will write $\alpha \vartriangleright\beta$ to denote that the following condition is satisfied: if $\g\neq\lo$, then $\q=\ql$ and:
\begin{enumerate}
 \item there exists $i\in\{1,...,n\}$ such that $\g_i = \g$
 \item for every $i=1,...,n$,  if $\g_i = \g$,  then  $\q_i=\ql$
\end{enumerate}

Let $pro=(\St,\e)$ be a program of $L^1[\Sigma^\q]$ (resp. $L^1[\Sigma^\g]$), and let $\Sigma^\q_{pro}$ (resp. $\Sigma^\g_{pro}$) be the list of operators and constructors (with their types), listed in order of occurrence. In section \ref{glosariodecasosdeestudio} we show $pro$ and the lists $\Sigma^\q_{pro},\ \Sigma^\g_{pro}$ for several case studies.
We will write $\Sigma^\q_{pro} \vartriangleright\Sigma^\g_{pro}$ to denote that there exists a one-to-one correspondence between $\Sigma^\q_{pro}$ and $\Sigma^\g_{pro}$ that maps every operator $\op^\tau$ into an operator $\op^\theta$ satisfying $\tau\vartriangleright\theta$, and every constructor $\ct^\sigma$ into a constructor $\ct^\rho$ satisfying $\sigma\vartriangleright\rho$.
Note that by definition, neither the 0-ary operators nor the constructor $[]^\sigma$ can have a global type of the form $\x\ \Pt$ as output without breaking the property $\Sigma^\q_{pro} \vartriangleright\Sigma^\g_{pro}$.

The problem of obtaining qualifications $\Sigma^\q_{pro}$ (resp. $\Sigma^\g_{pro}$) that grant linearity (resp. globality) to an unqualified program is a non-trivial problem, which requires solving the problem of an exponential search. The linearizations (resp. globalizations)  shown in section  \ref{glosariodecasosdeestudio} are the most efficient in the list returned by the algorithm of section  \ref{algoritmodelinealizacion} (resp. section  \ref{algoritmodeglobalizacion}).

To illustrate our case studies we will show the program $pro=(\St,\e)$ with unqualified operators and constructors, along with the lists $\Sigma^\q_{pro}$ and $\Sigma^\g_{pro}$ that give linearity (resp. globality) to $pro$. We denote by $pro^{\Sigma^\q_{pro}}$ (resp. $pro^{\Sigma^\q_{pro}}$) the corresponding program of $L^1[\Sigma^\q]$ (resp. $L^1[\Sigma^\g]$.
We will use $pro^{\Sigma^\qu}$ for the pure functional program, that is, the program in which all qualifiers are $\qu$.

By $C\ P$ we denote the size of the memory used in the evaluation of $P$. We count the auxiliary variables generated in the evaluation by the function $new\ \E$. Boolean type variables are not counted, while integer type variables have size 1, and array type variables have size corresponding to their length.

For example, $C fib_1^{\Sigma^\qu_{fib_1}}=2n+3$. We can inductively deduce this formula. If $n=0$, the evaluation uses 3 variables, which correspond to the evaluation of $n,1,1$ respectively (we don't count the evaluation of $(==0)$ because it produces a boolean). Suppose the evaluation with $\x=n-1$ uses $2(n-1)+3$ variables. So the evaluation for $\x=n$ uses $2(n-1)+3 + 2= 2n+3$ variables (the last 2 corresponds to $(-1),(+)$).
The calculation of $C fib_1^{\Sigma^\q_{fib_1}}=n+3$ takes into account not only the variables that are created, but also those that are destroyed. In the evaluation of $(-1)$  one variable is destroyed.

The \textit{linear memory ratio} is defined by:
\[R_{pro}^{\Sigma^\q_{pro}}=\lim_{n\rightarrow\infty} \frac{C\ pro^{\Sigma^\q_{pro}}}{C\ pro^{\Sigma^\qu_{pro}}}\]
Similarly we define the \textit{global memory ratio}
$R_{pro}^{\Sigma^\g_{pro}}$.

Finally, the \textit{functional memory residue} is defined by:
\[r_{pro}^{\Sigma^\g_{pro},\Sigma^\q_{pro}}=C\ pro^{\Sigma^\g_{pro}}-C\ pro^{\Sigma^\q_{pro}}\]
This last defined magnitude represents how much of the improvement of the linear program with respect to the unrestricted one is capitalized by the imperative version.

We will use the following terms to evaluate each case study $pro,\Sigma^\q_{pro},\Sigma^\g_{pro}$:
\begin{itemize}
 \item \textit{Protected}: $\Sigma^\q_{pro}\triangleright\Sigma^\g_{pro}$.
 \item \textit{Linear improvement}: $R_{pro}^{\Sigma^\q_{pro}}=0$.
 \item \textit{Imperative improvement}: $R_{pro}^{\Sigma^\g_{pro}}=0$.
\item \textit{Full  linear}: $C\ pro^{\Sigma^\q_{pro}}$ does not depend on $n$.
\item \textit{Full  imperative}: $C\ pro^{\Sigma^\g_{pro}}$ does not depend on $n$.
 \item \textit{Linear-Imperative match}: $r_{pro}^{\Sigma^\g_{pro},\Sigma^\q_{pro}}$ does not depend on $n$. For short, we write \textit{LI-match}.
\end{itemize}

For list algorithms, we take the polynomial $C\ pro^{\Sigma^\q_{pro}}-2n$ to evaluate the membership of $pro$ to the different categories, since $2n$ is the cost of generating a list of size $n$.

In section \ref{glosariodecasosdeestudio}, a sequence of transformations\footnote{This consists of increasing the number of parameters of the functions and using the operators $\id$ and $\pi_i$ to achieve protection. }
can be seen for each algorithm that aims to achieve a protected full  version (that is, protected, full linear and full imperative).
In this path, there are intermediate attempts in which the maximum memory saving is not achieved ($fib_1$, $fact_2$), or it is achieved in some of the versions but there is no LI-match ($map_1$, $fib_4$). In other cases, LI-match is achieved but not protection ($fib_2$, $map_2$, $insl_1$).

It can be observed in the protected full cases ($fib_5$, $fact_3$, $map_3$ and $insl_2$), which we could consider as the most successful, that the syntax of the original functional program is significantly altered to force LI-match, that is, to force the secure in-place update to be adequately modeled by the linearity property.

\subsection{Forcing protection}
\label{forcingprotection}

Linear operators/constructors can be replaced by operators/constructors with a higher level of destructiveness, to increase their protection capacity. This is done by adding an input that will only fulfill the role of being destroyed in the linear evaluation.
In this way, linear evaluation can model the destruction that the in-place update will perform, and thus ensure that this operation will not be harmful. The semantics of these new operators/constructors are identical to the semantics of the original, except that in linear evaluation, the destruction of the added input is performed. This resource not only allows us to increase the protection capacity, but also allows us to obtain new LI-match situations.

For example, the operation $\cdot[\cdot, \cdot] : (\qh\ \Array,\qh\ \Int,\ql\ \Int)\rightarrow\ql\ \Int$ of $map_3$ is a case of forced protection. The addition of the third input allows us to obtain protection for $\cdot[\cdot, \cdot] : (\ara\ \Array,\vi\ \Int,\z\ \Int)\rightarrow\z\ \Int$. Simultaneously, we obtain LI-match, thus making the algorithm protected full.

The constructor $[:] : (\ql\ [\ql\ \Int],\ql\ \Int,\ql\ [\ql\ \Int])\rightarrow\ql\ [\ql\ \Int]$ presents another example of forced protection. The linear evaluation of $[\xs](\z:\zs)$ yields the same result as $(\z:\zs)$, and also destroys the memory location $\xs$. Because of the linear typing that enables this destruction, the in-place update $\xs:=(\z:\zs)$ is protected. Again, we have a  protected full algorithm.

\subsection{Commands with global variables (only)}
\label{comandosconvariablesglobales}

Full programs, where all variables that occur in the types are free variables, produces commands in which no parameter passing takes place. In this case, global variables are not passed as parameters: the transformation $P$ given in section \ref{formaimperativa} excludes them from the formal parameters of functions and $\Let$ statements (we use $\pp$, which denotes the pattern $\p$ in which variables in $Gl\ \Gamma$ are replaced by $\ep$).

To reflect this fact in the small-step semantics of $L^1[\Sigma^\g]$ we must replace the relation $(\St,\e)\rightarrow(\St',\e')$ by the relation $(\St,\e)\rightarrow^G(\St',\e')$, where $G$ is a set of variables. The rule ($\mathsf{ele})$ must take the following form:
\medskip

$(\St, \Let\ \p\equiv\ \p'\ \In\  \e) \rightarrow^G_\beta (\St, [\pp^G\mapsto\p']\e) $

\medskip

\noindent 
Here $\pp^G$ denotes the pattern $\p$ in which variables of $G$ are replaced by $\ep$ (note that $[\ep\mapsto\p_0]$ is the trivial substitution $[]$). A similar modification must be made for the rule ($\mathsf{eap}$). The rule ($\mathsf{eca}$) does not need to be modified since global typing prevents parameters of the phrase $\Case$ from being global.

Examples of this kind of programs are $fib_{5\g}$ and $fact_{5\g}$. In this last case it should be noted that the imperative program consists of a simple cycle of the style $\mathsf{while}\ \arb\ \Do\ \arc$ (see definition of this phrase in the theoretical language Iswim \cite{reynolds}). The same is observed if we obtain $\Map_{2\g}$ from $\Map_2$, declaring $\vi,\n,\z$ global. The last two lines of the program $P\ \Gamma^\g\ \Map_{2g}^{}$ become:

\smallskip
$\mathsf{while}\ \neg (\vi==\n)\  \mathsf{do}\ \z:=\ara[\vi];\  \mathsf{fun}();\  \ara[\vi]:=\z;\  \vi:=(+1)\ \vi$.

\section{Conclusions}
\label{conclusiones}

Computational interpretations of linear logic provide an improvement in the use of memory resources, and suggest a theoretical model for the safe introduction of in-place update. In these approaches, a type system ``protects'' the imperative version by preventing harmful destructive use of memory.
In this work we establish a conceptually clear relationship between the two languages, and address the task of ``measuring'' the improvement given by a linear program, and how much of it is preserved in the imperative version.
The concepts of linear ratio and functional residue formalize these measures, and the categories optimal and full  allow us to classify a good number of case studies. Both the efficiency and the ability of linearity to model the in-place update are adequately reflected in these categories.

The key to the work is the definition of a language in which the qualified signature defines three different languages: functional or unrestricted ($L^1[\Sigma^\qu]$), linear ($L^1[\Sigma^\q]$) and global ($L^1[\Sigma^\g]$), which introduces in-place update and global variables. Possibly the theoretical model of introduction of imperative elements given by $L^1[\Sigma^\g]$ and the transformation $P\ \Gamma^\g\ pro^{\Sigma^\g}$ are the main contributions of this work. The strategy of condensing the imperative attributes of the language in the signature provides theoretical transpariency, follows the spirit of  Iswim language  \cite{reynolds} in the integration of the two paradigms.

The relation $\Sigma_{pro}^\q\vartriangleright\Sigma_{pro}^\g$ states that linear evaluation ``protects'' imperative execution. If the program is linearly well-typed (which guarantees its linear evaluation), the correctness of the associated imperative program will be guaranteed.

The strategy of putting all types (integers, booleans, arrays) at the same level, even when they have different memory requirements, marks the intention of establishing a general theoretical framework in which to study the problem. For simplicity we have incorporated only one recursive type (list), but clearly the language could be extended with other recursive types without much difficulty.

We consider the results shown in the case studies to be encouraging. The "more substructural" of the linearizations protects at least one well-typed globalization (generated by the algorithm). But the global type checking algorithm needs to be improved.
Since global typing rules are strongly nondeterministic, the type checking algorithm resolves nondeterminism by forcing globality (for example, in rule \textsf{bog}, modified at the end of the \ref{sistemadetipoglobal} section). This strategy corresponds to the strategy of forcing a global qualifier for operator inputs when finding globalizations (set $C_i$ in section \ref{algoritmodeglobalizacion}), to increase the chances that the operator can be protected. As a consequence of these strategies, the global type checking algorithm is too restrictive.
For example, the program $\x+(\Let\ \z\equiv\y\ \In\ \z)$, which admits trivial globalization (all qualifiers $\lo$), is rejected by the type checking algorithm, since it forces global type ($\y\ \Int$) on the second argument of the sum. Note that $p^\Gamma(\Let\ \z\equiv\y\ \In\ \z)=\y$, even though $\Gamma\y=\lo\ \Int$.

In general, the incorporation of imperative elements into functional programs is a complex process, which still requires the development of more general theoretical frameworks. This work intends to be a contribution in this direction.

\section{Case study glossary}
\label{glosariodecasosdeestudio}

All programs and results in this section were obtained with a  prototyte\footnote{https://github.com/hgramaglia/L1} developed in Haskell.

To show the program $pro=(\St,\e)$ we arrange the definitions of $\St$ and the expression $\e$ in successive lines (without the parentheses).
In this section we remove the types $\tau$ and $\sigma$ from the operations $\op^\tau$ and the constructors $[]^\sigma$, $(:)^\sigma$. We show the lists $\Sigma^\q_{pro}$ and $\Sigma^\g_{pro}$ considering the operations and constructors in the order of occurrence in $(\St,\e)$. We also show the type-qualified environments $\Gamma^\q$ and $\Gamma^\g$. In each case, following the store, the expression, and the respective lists, we also show the imperative form given by $P\ \Gamma^\g\ pro^{\Sigma^\g_{pro}}$.
If $FV\ \p=\emptyset$, then we use the abbreviation $(\Let\ \p\equiv\e\ \In\ \e')=_{def} \e; \e'$. (see \cite{reynolds}).

A brief description is added to each case study, analyzing the performance according to the categories defined in section  \ref{resumendecasosdeestudio}, also indicating whether the protection property $\Sigma^\q_{pro}\triangleright\Sigma^\g_{pro}$ is verified (although in reality it is very easy to check from the two signatures shown in parallel). Except for the counterexample $case$, all algorithms are well-typed, both linearly and globally.

We will often use unary operators (for example $(==\!\!0)\ \x$ or $(-\!1)\ \x$) instead of their usual binary forms ($\x==0$ or $\x-1$, respectively). This change is essential in many cases to achieve the LI-match property.

Finally, to save space, in some cases we write $\Sigma^\q_{-}$ instead of $\Sigma^\q_{pro}$.

\bigskip

\noindent$fib_1$
$
\tiny\begin{array}{|ll}
\hline
\\
\Fib\ = \ \lambda \x .\ \If\ (==0)\ \x \ \Then\ \La\x,1,1\Ra\ \Else\ \Let\ \La\x,\w,\y\Ra \equiv \Fib\ ((-1)\ \x) \  \In\ \La\x,\y,(\w+\y)\Ra,\\
\Fib\ n\\
\textsf{Protected LI-match. Since linear memory ratio is $1/2$, it is neither linear improvement nor global improvement.}\\
\end{array}
$

\noindent\begin{tabular}{|c |c |}
\hline
$
\begin{array}{l}

\tiny\begin{array}{lcll}
\\
\Sigma^\q_{fib_1}
& ==0 & : & \qh\ \Int\rightarrow\ql\ \Bool\\
& 1 & : & \qu\ \Int\\
& 1 & : & \qu\ \Int\\
& -1 & : & \ql\ \Int\rightarrow\ql\ \Int\\
& + & : & (\qu\ \Int,\qu\ \Int)\rightarrow\qu\ \Int\\
& n & : & \ql\ \Int\smallskip\\
\end{array}
\\
\tiny\begin{array}{llll}
\Gamma^\q
& \Fib & : & \ql\ \Int\rightarrow\La\ql\ \Int,\qu\ \Int,\qu\ \Int\Ra\\
\end{array}

\end{array}
$
&
$
\begin{array}{l}

\tiny\begin{array}{lcll}
\Sigma^\g_{fib_1}
& ==0 & : & \x\ \Int\rightarrow\lo\ \Bool\\
& 1 & : & \lo\ \Int\\
& 1 & : & \lo\ \Int\\
& -1 & : & \x\ \Int\rightarrow\x\ \Int\\
& + & : & (\lo\ \Int,\lo\ \Int)\rightarrow\lo\ \Int\\
& n & : & \lo\ \Int\smallskip\\
\end{array}
\\
\tiny\begin{array}{llll}
\Gamma^\g
& \Fib & : & \Pi\x:\lo\ \Int.\ \La\x\ \Int,\lo\ \Int,\lo\ \Int\Ra\\
\end{array}

\end{array}
$\\
\hline
\end{tabular}\\

\noindent$\tiny\begin{array}{|ll}
\Fib\ = \ \lambda \x .\ \If\ (==0)\ \x \ \Then\ \La\x,1,1\Ra\ \Else\ \Let\ \La\x,\w,\y\Ra \equiv \Fib\ (\x := (-1)\ \x) \  \In\ \La\x,\y,(\w+\y)\Ra,\\
\Fib\ n\\
\hline
\end{array}
$
$P\ \Gamma^\g\ fib_1^{\Sigma^\g_{fib_1}}$

\bigskip

\noindent$fib_2$
$
\tiny\begin{array}{|ll}
\hline
\Fib\ = \ \lambda \x .\ \If\ (==0)\ \x \ \Then\ \La\x,1,1\Ra\ \Else\ \Let\ \La\x,\w,\y\Ra \equiv \Fib\ ((-1)\ \x) \  \In\ \La\x,\id(\y),(\w+\y)\Ra,\\
\Fib\ n\\
\textsf{Full not protected ($\id:\y\ \Int\rightarrow\w\ \Int$ has no protection). The global version is incorrect.}\\
\textsf{
The operator $\id$ is used to achieve correct linear qualification (see \cite{gramagliawlt}).}\\
\end{array}
$

\noindent\begin{tabular}{|c |c |}
\hline
$
\begin{array}{l}
\tiny\begin{array}{lcll}
\Sigma^\q_{fib_2}
& ==0 & : & \qh\ \Int\rightarrow\ql\ \Bool\\
& 1 & : & \ql\ \Int\\
& 1 & : & \ql\ \Int\\
& -1 & : & \ql\ \Int\rightarrow\ql\ \Int\\
& \id & : & \qh\ \Int\rightarrow\ql\ \Int\\
& + & : & (\ql\ \Int,\ql\ \Int)\rightarrow\ql\ \Int\qquad\\
& n & : & \ql\ \Int\smallskip\\
\end{array}
\\
\tiny\begin{array}{llll}
\Gamma^\q
& \Fib & : & \ql\ \Int\rightarrow\La\ql\ \Int,\ql\ \Int,\ql\ \Int\Ra\\
\end{array}
\end{array}
$
&
$
\begin{array}{l}
\tiny\begin{array}{lcll}
\Sigma^\g_{fib_2}
& ==0 & : & \x\ \Int\rightarrow\lo\ \Bool\\
& 1 & : & \w\ \Int\\
& 1 & : & \y\ \Int\\
& -1 & : & \x\ \Int\rightarrow\x\ \Int\\
& \id & : & \y\ \Int\rightarrow\w\ \Int\\
& + & : & (\w\ \Int,\y\ \Int)\rightarrow\y\ \Int\qquad\\
& n & : & \lo\ \Int\smallskip\\
\end{array}
\\
\tiny\begin{array}{llll}
\Gamma^\g
& \Fib & : & \Pi\x:\lo\ \Int.\ \La\x\ \Int,\w\ \Int,\y\ \Int\Ra\\
\end{array}
\end{array}
$\\
\hline
\end{tabular}\\
$\tiny\begin{array}{|ll}
\Fib\ = \ \lambda \x .\ \If\ (==0)\ \x \ \Then\ \La\x,\w := 1,\y := 1\Ra\\
\qquad\qquad\qquad\qquad\quad\Else\ \Let\ \La\x,\w,\y\Ra \equiv \Fib\ (\x := (-1)\ \x) \  \In\ \La\x,\w := \y,\y := (\w+\y)\Ra,\\
\Fib\ n\\
\hline
\end{array}
$
$P\ \Gamma^\g\ fib_2^{\Sigma^\g_{fib_2}}$

\bigskip

\noindent$fib_3$
$
\tiny\begin{array}{|ll}
\hline
\Fib\ = \ \lambda \La\x,\w,\y\Ra .\ \If\ (==0)\ \x \ \Then\ \La\x,\w,\y\Ra\ \Else\ \Let\ \La\x,\w,\y\Ra \equiv \Fib\ \La(-1)\ \x,\w,\y\Ra \  \In\ \La\x,\pi_2(\w,\y),(\w+\y)\Ra,\\
\Fib\ \La n,1,1\Ra\\
\textsf{Full sin protección. Intenta aumentar las chances de correcta globalización de $fib_2$: la proyección $\pi_2(\w,\y)$ se utiliza como versión}\\
\textsf{global de $\id(\y)$. Sigue sin protección: $w$ tiene tipo $\qh\ \Int$ como argumento de $\pi_2$. Cómo $fib2$, el programa imperativo resultante}\\
\textsf{no es correcto.}\\
\end{array}
$

\noindent\begin{tabular}{|c |c |}
\hline
$
\begin{array}{l}

\tiny\begin{array}{lcll}
\Sigma^\q_{fib_3}
& ==0 & : & \qh\ \Int\rightarrow\ql\ \Bool\\
& -1 & : & \ql\ \Int\rightarrow\ql\ \Int\\
& \pi_2 & : & (\qh\ \Int,\qh\ \Int)\rightarrow\ql\ \Int\\
& + & : & (\ql\ \Int,\ql\ \Int)\rightarrow\ql\ \Int\\
& n & : & \ql\ \Int\\
& 1 & : & \ql\ \Int\\
& 1 & : & \ql\ \Int\smallskip\\
\end{array}
\\
\tiny\begin{array}{llll}
\Gamma^\q
& \Fib & : & \La\ql\ \Int,\ql\ \Int,\ql\ \Int\Ra\rightarrow\La\ql\ \Int,\ql\ \Int,\ql\ \Int\Ra\\
\end{array}

\end{array}
$
&
$
\begin{array}{l}

\tiny\begin{array}{lcll}
\Sigma^\g_{fib_3}
& ==0 & : & \x\ \Int\rightarrow\lo\ \Bool\\
& -1 & : & \x\ \Int\rightarrow\x\ \Int\\
& \pi_2 & : & (\w\ \Int,\y\ \Int)\rightarrow\w\ \Int\\
& + & : & (\w\ \Int,\y\ \Int)\rightarrow\y\ \Int\\
& n & : & \lo\ \Int\\
& 1 & : & \lo\ \Int\\
& 1 & : & \lo\ \Int\smallskip\\
\end{array}
\\
\tiny\begin{array}{llll}
\Gamma^\g
& \Fib & : & \Pi\La\x,\w,\y\Ra:\La\lo\ \Int,\lo\ \Int,\lo\ \Int\Ra.\\
& & &\qquad \qquad \La\x\ \Int,\w\ \Int,\y\ \Int\Ra\\
\end{array}

\end{array}
$\\
\hline
\end{tabular}\\
$\tiny\begin{array}{|ll}
\Fib\ = \ \lambda \La\x,\w,\y\Ra .\ \If\ (==0)\ \x \ \Then\ \La\x,\w,\y\Ra\ \\
\qquad\qquad\qquad\qquad\qquad\quad\Else\ \Let\ \La\x,\w,\y\Ra \equiv \Fib\ \La\x := (-1)\ \x,\w,\y\Ra \  \In\ \La\x,\w := \y,\y := (\w+\y)\Ra,\\
\Fib\ \La n,1,1\Ra\\
\hline
\end{array}
$
$P\ \Gamma^\g\ fib_3^{\Sigma^\g_{fib_3}}$

\bigskip

\noindent$fib_4$
$
\tiny\begin{array}{|ll}
\hline
\Fib\ = \ \lambda \La\x,\w,\y\Ra .\ \If\ (==0)\ \x \ \Then\ \La\x,\w,\y\Ra\\
\qquad\qquad\qquad\qquad\qquad\qquad \Else\ \Let\ \La\x,\w,\y\Ra \equiv \Fib\ \La(-1)\ \x,\w,\y\Ra \  \In\ \Let\ \z \equiv \id(\w) \  \In\ \La\x,\pi_2(\w,\y),(\z+\y)\Ra,\\
\Fib\ \La n,1,1\Ra\\
\textsf{Protected full linear. It is neither global improvement nor LI-match, but the global version improves the unrestricted:}\\
\textsf{global memory ratio is $1/4$. The $\id$ operator is used to achieve correct qualification (see \cite{gramagliawlt}).}\\
\end{array}
$

\noindent\begin{tabular}{|c |c |}
\hline
$
\begin{array}{l}
\tiny\begin{array}{lcll}
\Sigma^\q_{fib_4}
& ==0 & : & \qh\ \Int\rightarrow\ql\ \Bool\\
& -1 & : & \ql\ \Int\rightarrow\ql\ \Int\\
& \id & : & \qh\ \Int\rightarrow\ql\ \Int\\
& \pi_2 & : & (\ql\ \Int,\qh\ \Int)\rightarrow\ql\ \Int\\
& + & : & (\ql\ \Int,\ql\ \Int)\rightarrow\ql\ \Int\\
& n & : & \ql\ \Int\\
& 1 & : & \ql\ \Int\\
& 1 & : & \ql\ \Int\smallskip\\
\end{array}
\\
\tiny\begin{array}{llll}
\Gamma^\q
& \Fib & : & \La\ql\ \Int,\ql\ \Int,\ql\ \Int\Ra\rightarrow\La\ql\ \Int,\ql\ \Int,\ql\ \Int\Ra\\
\end{array}
\end{array}
$
&
$
\begin{array}{l}
\tiny\begin{array}{lcll}
\Sigma^\g_{fib_4}
& ==0 & : & \x\ \Int\rightarrow\lo\ \Bool\\
& -1 & : & \x\ \Int\rightarrow\x\ \Int\\
& \id & : & \w\ \Int\rightarrow\lo\ \Int\\
& \pi_2 & : & (\w\ \Int,\y\ \Int)\rightarrow\w\ \Int\\
& + & : & (\lo\ \Int,\y\ \Int)\rightarrow\y\ \Int\\
& n & : & \lo\ \Int\\
& 1 & : & \lo\ \Int\\
& 1 & : & \lo\ \Int\smallskip\\
\end{array}
\\
\tiny\begin{array}{llll}
\Gamma^\g
& \Fib & : & \Pi\La\x,\w,\y\Ra:\La\lo\ \Int,\lo\ \Int,\lo\ \Int\Ra.\\
& & & \qquad\La\x\ \Int,\w\ \Int,\y\ \Int\Ra\\
\end{array}
\end{array}
$\\
\hline
\end{tabular}\\
$\tiny\begin{array}{|ll}
\Fib\ = \ \lambda \La\x,\w,\y\Ra .\ \If\ (==0)\ \x \ \Then\ \La\x,\w,\y\Ra\\
\qquad\qquad\qquad\quad\Else\ \Let\ \La\x,\w,\y\Ra \equiv \Fib\ \La\x := (-1)\ \x,\w,\y\Ra \  \In\ \Let\ \z \equiv \w \  \In\ \La\x,\w := \y,\y := (\z+\y)\Ra,\\
\Fib\ \La n,1,1\Ra\\
\hline
\end{array}
$
$P\ \Gamma^\g\ fib_4^{\Sigma^\g_{fib_4}}$

\bigskip

\noindent$fib_5$
$
\tiny\begin{array}{|ll}
\hline
\Fib\ = \ \lambda \La\x,\w,\y,\z\Ra .\ \If\ (==0)\ \x \ \Then\ \La\x,\w,\y,\z\Ra\\
\qquad\qquad\qquad\qquad \Else\ \Let\ \La\x,\w,\y,\z\Ra \equiv \Fib\ \La(-1)\ \x,\w,\y,\z\Ra \  \In\ \Let\ \z \equiv \pi_2(\z,\w) \  \In\ \La\x,\pi_2(\w,\y),(\z+\y),\z\Ra,\\
\Fib\ \La n,1,1,1\Ra\\
\textsf{Protected full.}\\
\end{array}
$

\noindent\begin{tabular}{|c |c |}
\hline
$
\begin{array}{l}

\tiny\begin{array}{lcll}
\Sigma^\q_{fib_5}
& ==0 & : & \qh\ \Int\rightarrow\ql\ \Bool\\
& -1 & : & \ql\ \Int\rightarrow\ql\ \Int\\
& \pi_2 & : & (\ql\ \Int,\qh\ \Int)\rightarrow\ql\ \Int\\
& \pi_2 & : & (\ql\ \Int,\qh\ \Int)\rightarrow\ql\ \Int\\
& + & : & (\qh\ \Int,\ql\ \Int)\rightarrow\ql\ \Int\\
& n & : & \ql\ \Int\\
& 1 & : & \ql\ \Int\\
& 1 & : & \ql\ \Int\\
& 1 & : & \ql\ \Int\smallskip\\
\end{array}
\\
\tiny\begin{array}{llll}
\Gamma^\q
& \Fib & : & \La\ql\ \Int,\ql\ \Int,\ql\ \Int,\ql\\
& & & \ \Int\Ra\rightarrow\La\ql\ \Int,\ql\ \Int,\ql\ \Int,\ql\ \Int\Ra\\
\end{array}
\end{array}
$
&
$
\begin{array}{l}
\tiny\begin{array}{lcll}
\Sigma^\g_{fib_5}
& ==0 & : & \x\ \Int\rightarrow\lo\ \Bool\\
& -1 & : & \x\ \Int\rightarrow\x\ \Int\\
& \pi_2 & : & (\z\ \Int,\w\ \Int)\rightarrow\z\ \Int\\
& \pi_2 & : & (\w\ \Int,\y\ \Int)\rightarrow\w\ \Int\\
& + & : & (\z\ \Int,\y\ \Int)\rightarrow\y\ \Int\\
& n & : & \lo\ \Int\\
& 1 & : & \lo\ \Int\\
& 1 & : & \lo\ \Int\\
& 1 & : & \lo\ \Int\smallskip\\
\end{array}
\\
\tiny\begin{array}{llll}
\Gamma^\g
& \Fib & : & \Pi\La\x,\w,\y,\z\Ra:\La\lo\ \Int,\lo\ \Int,\lo\ \Int,\lo\ \Int\Ra.\\
& & &\
\qquad\qquad\La\x\ \Int,\w\ \Int,\y\ \Int,\z\ \Int\Ra\\
\end{array}

\end{array}
$\\
\hline
\end{tabular}\\
$\tiny\begin{array}{|ll}
\Fib\ = \ \lambda \La\x,\w,\y,\z\Ra .\ \If\ (==0)\ \x \ \Then\ \La\x,\w,\y,\z\Ra\\
\qquad\qquad\qquad\qquad \Else\ \Let\ \La\x,\w,\y,\z\Ra \equiv \Fib\ \La\x := (-1)\ \x,\w,\y,\z\Ra \  \In\\
\qquad\qquad\qquad\qquad\quad\ \  \Let\ \z \equiv (\z := \w) \  \In\ \La\x,\w := \y,\y := (\z+\y),\z\Ra,
\qquad\qquad\qquad\qquad \\
\Fib\ \La n,1,1,1\Ra\\
\hline
\end{array}
$
$P\ \Gamma^\g\ fib_5^{\Sigma^\g_{fib_5}}$

\bigskip

\noindent$fact_1$
$
\tiny\begin{array}{|ll}
\hline
\Fact\ = \ \lambda \x .\ \If\ (==0)\ \x \ \Then\ 1\ \Else\ (\x*\Fact\ ((-1)\ \x)),\\
\Fact\ 10\\
\textsf{
LI-match not protected. It is neither linear improvement nor imperative improvement: the memory ratio is $1/2$. }\\
\textsf{
For the globalization algorithm to work, w must be added to the type context, as it occurs free in the type of the expression.}\\
\end{array}
$

\noindent\begin{tabular}{|c |c |}
\hline
$
\begin{array}{l}
\tiny\begin{array}{lcll}
\Sigma^\q_{fact_1}
& ==0 & : & \qu\ \Int\rightarrow\ql\ \Bool\\
& 1 & : & \ql\ \Int\\
& * & : & (\qu\ \Int,\ql\ \Int)\rightarrow\ql\ \Int\\
& -1 & : & \qu\ \Int\rightarrow\qu\ \Int\\
& n & : & \qu\ \Int\smallskip\\
\end{array}
\\
\tiny\begin{array}{llll}
\Gamma^\q
& \Fact & : & \qu\ \Int\rightarrow\ql\ \Int\\
\end{array}
\end{array}
$
&
$
\begin{array}{l}
\tiny\begin{array}{lcll}
\Sigma^\g_{fact_1}
& ==0 & : & \lo\ \Int\rightarrow\lo\ \Bool\\
& 1 & : & \w\ \Int\\
& * & : & (\lo\ \Int,\w\ \Int)\rightarrow\w\ \Int\\
& -1 & : & \lo\ \Int\rightarrow\lo\ \Int\\
& n & : & \lo\ \Int\smallskip\\
\end{array}
\\
\tiny\begin{array}{llll}
\Gamma^\g & \w & : & \w\ \Int\\
& \Fact & : & \Pi\La\Ra:\lo\ \Int.\ \w\ \Int\\
\end{array}
\end{array}
$\\
\hline
\end{tabular}\\ 
$\tiny\begin{array}{|ll}
\w=1,\ \Fact\ = \ \lambda \x .\ \If\ (==0)\ \x \ \Then\ \w := 1\ \Else\ \w := (\x*\Fact\ ((-1)\ \x)),\\
\Fact\ n\\
\hline
\end{array}
$
$P\ \Gamma^\g\ fact_1^{\Sigma^\g_{fact_1}}$

\bigskip

\noindent$fact_2$
$
\tiny\begin{array}{|ll}
\hline
\Fact\ = \ \lambda \La\x,\w\Ra .\ \If\ (==0)\ \x \ \Then\ \w\ \Else\ (\x*\Fact\ \La(-1)\ \x,\w\Ra),\\
\Fact\ \La n,1\Ra\\
\textsf{Protected LI-match, but it is neither linear improvement nor imperative improvement: the memory ratio is $1/2$.}\\
\end{array}
$

\noindent\begin{tabular}{|c |c |}
\hline
$
\begin{array}{l}

\tiny\begin{array}{lcll}
\Sigma^\q_{fact_2}
& ==0 & : & \qu\ \Int\rightarrow\ql\ \Bool\\
& * & : & (\qu\ \Int,\ql\ \Int)\rightarrow\ql\ \Int\\
& -1 & : & \qu\ \Int\rightarrow\qu\ \Int\\
& n & : & \qu\ \Int\\
& 1 & : & \ql\ \Int\smallskip\\
\end{array}
\\
\tiny\begin{array}{llll}
\Gamma^\q
& \Fact & : & \La\qu\ \Int,\ql\ \Int\Ra\rightarrow\ql\ \Int\\
\end{array}

\end{array}
$
&
$
\begin{array}{l}

\tiny\begin{array}{lcll}
\Sigma^\g_{fact_2}
& ==0 & : & \lo\ \Int\rightarrow\lo\ \Bool\\
& * & : & (\lo\ \Int,\w\ \Int)\rightarrow\w\ \Int\\
& -1 & : & \lo\ \Int\rightarrow\lo\ \Int\\
& n & : & \lo\ \Int\\
& 1 & : & \lo\ \Int\smallskip\\
\end{array}
\\
\tiny\begin{array}{llll}
\Gamma^\g
& \Fact & : & \Pi\La\x,\w\Ra:\La\lo\ \Int,\lo\ \Int\Ra.\ \w\ \Int\\
\end{array}

\end{array}
$\\
\hline
\end{tabular}\\ 
$\tiny\begin{array}{|ll}
\Fact\ = \ \lambda \La\x,\w\Ra .\ \If\ (==0)\ \x \ \Then\ \w\ \Else\ \w := (\x*\Fact\ \La(-1)\ \x,\w\Ra),\\
\Fact\ \La n,1\Ra\\
\hline
\end{array}
$
$P\ \Gamma^\g\ fact_2^{\Sigma^\g_{fact_2}}$

\bigskip

\noindent$fact_3$
$
\tiny\begin{array}{|ll}
\hline
\Fact\ = \ \lambda \La\w,\x\Ra .\ \If\ (==0)\ \x \ \Then\ \La\w,\x\Ra\ \Else\ \Fact\ \La(\x*\w),(-1)\ \x\Ra,\\
\Fact\ \La1,n\Ra\\
\textsf{Protected full.}\\
\end{array}
$

\noindent\begin{tabular}{|c |c |}
\hline
$
\begin{array}{l}
\tiny\begin{array}{lcll}
\Sigma^\q_{fact_3}
& ==0 & : & \qh\ \Int\rightarrow\ql\ \Bool\\
& * & : & (\qh\ \Int,\ql\ \Int)\rightarrow\ql\ \Int\\
& -1 & : & \ql\ \Int\rightarrow\ql\ \Int\\
& 1 & : & \ql\ \Int\\
& n & : & \ql\ \Int\smallskip\\
\end{array}
\\
\tiny\begin{array}{llll}
\Gamma^\q
& \Fact & : & \La\ql\ \Int,\ql\ \Int\Ra\rightarrow\La\ql\ \Int,\ql\ \Int\Ra\\
\end{array}
\end{array}
$
&
$
\begin{array}{l}
\tiny\begin{array}{lcll}
\Sigma^\g_{fact_3}
& ==0 & : & \x\ \Int\rightarrow\lo\ \Bool\\
& * & : & (\x\ \Int,\w\ \Int)\rightarrow\w\ \Int\\
& -1 & : & \x\ \Int\rightarrow\x\ \Int\\
& 1 & : & \lo\ \Int\\
& n & : & \lo\ \Int\smallskip\\
\end{array}
\\
\tiny\begin{array}{llll}
\Gamma^\g
& \Fact & : & \Pi\La\w,\x\Ra:\La\lo\ \Int,\lo\ \Int\Ra.\ \La\w\ \Int,\x\ \Int\Ra\\
\end{array}
\end{array}
$\\
\hline
\end{tabular}\\ 
$\tiny\begin{array}{|ll}
\Fact\ = \ \lambda \La\w,\x\Ra .\ \If\ (==0)\ \x \ \Then\ \La\w,\x\Ra\ \Else\ \Fact\ \La\w := (\x*\w),\x := (-1)\ \x\Ra,\\
\Fact\ \La1,n\Ra\\
\hline
\end{array}
$
$P\ \Gamma^\g\ fact_3^{\Sigma^\g_{fact_3}}$

\bigskip

\noindent$map_1$
$
\tiny\begin{array}{|ll}
\hline
\ara\ =\ \left\lbrace 0,1,...,n\right\rbrace ,\\ \Fun\ =\ \lambda \x .\ (+1)\ \x,\\ \Map\ = \ \lambda \La\ara,\vi,\n\Ra .\ \If\ (\vi==\n) \ \Then\ \La\ara,\vi,\n\Ra\ \Else\ \Let\ \z \equiv \ara[\vi] \  \In\ \Map\ \La\ara[\vi\rightarrow\Fun\ \z],(+1)\ \vi,\n\Ra,\\
\Map\ \La\ara,0,n\Ra\\
\textsf{
Protected full linear but not LI-match. It is imperative improvement but not full.}\\
\textsf{
Defining $\cdot[\cdot]$ with global type $\z\ \Int$ we have LI-match but loses protection.}
\end{array}
$

\noindent\begin{tabular}{|c |c |}
\hline
$
\begin{array}{l}
\tiny\begin{array}{lcll}
\Sigma^\q_{-}
& +1 & : & \ql\ \Int\rightarrow\ql\ \Int\\
& == & : & (\qh\ \Int,\qh\ \Int)\rightarrow\ql\ \Bool\\
& \cdot[\ \cdot\ ] & : & (\qh\ \Array,\qh\ \Int)\rightarrow\ql\ \Int\\
& \cdot[\cdot\leftarrow\cdot ] & : & (\ql\ \Array,\qh\ \Int,\ql\ \Int)\rightarrow\ql\ \Array\\
& +1 & : & \ql\ \Int\rightarrow\ql\ \Int\\
& 0 & : & \ql\ \Int\\
& n & : & \ql\ \Int\smallskip\\
\end{array}
\\
\tiny\begin{array}{llll}
\Gamma^\q
& \ara & : &\ql\ \Array,\\
& \Fun & : &\ql\ \Int\rightarrow\ql\ \Int,\\
& \Map & : & \La\ql\ \Array,\ql\ \Int,\ql\ \Int\Ra\\
&  & &\qquad\qquad\rightarrow\La\ql\ \Array,\ql\ \Int,\ql\ \Int\Ra\\
\end{array}
\end{array}
$
&
$
\begin{array}{l}
\tiny\begin{array}{lcll}
\Sigma^\g_{-}
& +1 & : & \z\ \Int\rightarrow\z\ \Int\\
& == & : & (\vi\ \Int,\n\ \Int)\rightarrow\lo\ \Bool\\
& \cdot[\ \cdot\ ] & : & (\ara\ \Array,\vi\ \Int)\rightarrow\lo\ \Int\\
& \cdot[\cdot\leftarrow\cdot ] & : & (\ara\ \Array,\vi\ \Int,\lo\ \Int)\rightarrow\ara\ \Array\\
& +1 & : & \vi\ \Int\rightarrow\i\ \Int\\
& 0 & : & \lo\ \Int\\
& n & : & \lo\ \Int\smallskip\\
\end{array}
\\
\tiny\begin{array}{llll}
\Gamma^\g
& \ara & : &\ara\ \Array,\\
& \Fun & : &\Pi\z:\lo\ \Int.\ \z\ \Int,\\
& \Map & : & \Pi\La\ara,\vi,\n\Ra:\La\ara\ \Array,\lo\ \Int,\lo\ \Int\Ra.\\
&  & & \qquad\qquad\qquad \La\ara\ \Array,\vi\ \Int,\n\ \Int\Ra\\
\end{array}
\end{array}
$\\
\hline
\end{tabular}\\ 
$\tiny\begin{array}{|ll}
\ara\ =\ \left\lbrace 0,1,...,n\right\rbrace ,\\ \Fun\ =\ \lambda \z .\ \z := (+1)\ \z,\\ \Map\ = \ \lambda \La\La\Ra,\vi,\n\Ra .\ \If\ (\vi==\n) \ \Then\ \La\ara,\vi,\n\Ra\ \Else\ \Let\ \z \equiv \ara[\vi] \  \In\ \Map\ \La\ara := \ara[\vi\rightarrow\Fun\ \z],\vi := (+1)\ \vi,\n\Ra,\\
\Map\ \La\ara,0,n\Ra\\
\hline
\end{array}
$
$P\ \Gamma^\g\ map_1^{\Sigma^\g_{map_1}}$

\bigskip

\noindent$insl_1$
$
\tiny\begin{array}{|ll}
\hline
\\n\ =\ n,\ \x\ =\ 0,\\
\List\ =\ \lambda \La\n,\x\Ra .\ \If\ (==0)\ \n \ \Then\ \pi_1(\pi_1([],\n),\x)\ \Else\ (\id(\x):\List\ \La(-1)\ \n,(+1)\ \x\Ra),\\ \Ins\ = \ \lambda \La\x,\xs\Ra .\ \Case\  \ql\ \xs\ \Of\ ((\x:[]),(\z:\zs)\rightarrow\ \If\ (\x\leq\z) \ \Then\ (\x:(\z:\zs))\ \Else\ (\z:\Ins\ \La\x,\zs\Ra)),\\
\Ins\ \La n,\List\ \La\n,\x\Ra\Ra\\
\textsf{
Full not protected. List constructors are not protected in their occurrences in function $\Ins$.}\\
\end{array}
$

\noindent\begin{tabular}{|c |c |}
\hline
$
\begin{array}{l}
\tiny\begin{array}{lcll}
\Sigma^\q_{-}
& ==0 & : & \qh\ \Int\rightarrow\ql\ \Bool\\
& \pi_1 & : & (\ql\ [\ql\ \Int],\ql\ \Int)\rightarrow\ql\ [\ql\ \Int]\\
& \pi_1 & : & (\ql\ [\ql\ \Int],\ql\ \Int)\rightarrow\ql\ [\ql\ \Int]\\
& [] & : & \ql\ [\ql\ \Int]\\
& \id & : & \qh\ \Int\rightarrow\qu\ \Int\\
& : & : & (\qu\ \Int,\ql\ [\ql\ \Int])\rightarrow\ql\ [\ql\ \Int]\\
& -1 & : & \ql\ \Int\rightarrow\ql\ \Int\\
& +1 & : & \ql\ \Int\rightarrow\ql\ \Int\\
& : & : & (\ql\ \Int,\ql\ [\ql\ \Int])\rightarrow\ql\ [\ql\ \Int]\\
& [] & : & \ql\ [\ql\ \Int]\\
& \leq & : & (\qh\ \Int,\qh\ \Int)\rightarrow\ql\ \Bool\\
& : & : & (\ql\ \Int,\ql\ [\ql\ \Int])\rightarrow\ql\ [\ql\ \Int]\\
& : & : & (\ql\ \Int,\ql\ [\ql\ \Int])\rightarrow\ql\ [\ql\ \Int]\\
& : & : & (\ql\ \Int,\ql\ [\ql\ \Int])\rightarrow\ql\ [\ql\ \Int]\\
& n & : & \ql\ \Int\smallskip\\
\end{array}
\\
\tiny\begin{array}{llll}
\Gamma^\q
& \n & : &\ql\ \Int,\\
& \x & : &\ql\ \Int,\\
& \List & : &\La\ql\ \Int,\ql\ \Int\Ra\rightarrow\ql\ [\ql\ \Int],\\
& \Ins & : & \La\ql\ \Int,\ql\ [\ql\ \Int]\Ra\rightarrow\ql\ [\ql\ \Int]\\
\end{array}
\end{array}
$
&
$
\begin{array}{l}

\tiny\begin{array}{lcll}
\Sigma^\g_{-}
& ==0 & : & \n\ \Int\rightarrow\lo\ \Bool\\
& \pi_1 & : & (\lo\ [\lo\ \Int],\x\ \Int)\rightarrow\lo\ [\lo\ \Int]\\
& \pi_1 & : & (\lo\ [\lo\ \Int],\n\ \Int)\rightarrow\lo\ [\lo\ \Int]\\
& [] & : & \lo\ [\lo\ \Int]\\
& \id & : & \x\ \Int\rightarrow\lo\ \Int\\
& : & : & (\lo\ \Int,\lo\ [\lo\ \Int])\rightarrow\lo\ [\lo\ \Int]\\
& -1 & : & \n\ \Int\rightarrow\n\ \Int\\
& +1 & : & \x\ \Int\rightarrow\x\ \Int\\
& : & : & (\lo\ \Int,\lo\ [\lo\ \Int])\rightarrow\xs\ [\lo\ \Int]\\
& [] & : & \lo\ [\lo\ \Int]\\
& \leq & : & (\lo\ \Int,\lo\ \Int)\rightarrow\lo\ \Bool\\
& : & : & (\lo\ \Int,\lo\ [\lo\ \Int])\rightarrow\xs\ [\lo\ \Int]\\
& : & : & (\lo\ \Int,\lo\ [\lo\ \Int])\rightarrow\lo\ [\lo\ \Int]\\
& : & : & (\lo\ \Int,\lo\ [\lo\ \Int])\rightarrow\xs\ [\lo\ \Int]\\
& n & : & \lo\ \Int\smallskip\\
\end{array}
\\
\tiny\begin{array}{llll}
\Gamma^\g
& \n & : &\n\ \Int,\\
& \x & : &\x\ \Int,\\
& \List & : &\Pi\La\n,\x\Ra:\La\n\ \Int,\x\ \Int\Ra.\ \lo\ [\lo\ \Int],\\
& \Ins & : & \Pi\La\w,\xs\Ra:\La\lo\ \Int,\lo\ [\lo\ \Int]\Ra.\ \xs\ [\lo\ \Int]\\
\end{array}
\end{array}
$\\
\hline
\end{tabular}\\
$\tiny\begin{array}{|ll}
\\n\ =\ n,\ \x\ =\ 0,\\
\List\ =\ \lambda \La\La\Ra,\La\Ra\Ra .\ \If\ (==0)\ \n \ \Then\ []\ \Else\ (\x:\List\ \La\n := (-1)\ \n,\x := (+1)\ \x\Ra),\\ \Ins\ = \ \lambda \La\w,\xs\Ra .\ \Case\  \lo\ \xs\ \Of\\
\qquad\qquad\qquad(\xs := (\w:[]),(\z:\zs)\rightarrow\ \If\ (\w\leq\z) \ \Then\ \xs := (\w:(\z:\zs))\ \Else\ \xs := (\z:\Ins\ \La\w,\zs\Ra)),\\
\Ins\ \La \n,\List\ \La\n,\x\Ra\Ra\\
\hline
\end{array}
$
$P\ \Gamma^\g\ insl_1^{\Sigma^\g_{insl_1}}$

\bigskip

\noindent$mapl_1$
$
\tiny\begin{array}{|ll}
\hline
\n\ =\ n,\ \x\ =\ 0,\\
\F\ =\ \lambda \y .\ (*2)\ \y,\\ \List\ =\ \lambda \La\n,\x\Ra .\ \If\ (==0)\ \n \ \Then\ \pi_1(\pi_1([],\n),\x)\ \Else\ (\id(\x):\List\ \La(-1)\ \n,(+1)\ \x\Ra),\\ \Map\ = \ \lambda \La\f,\xs\Ra .\ \Case\  \ql\ \xs\ \Of\ ([],(\z:\zs)\rightarrow\ (\F\ \z:\Map\ \La\f,\zs\Ra)),\\
\Map\ \La\f,\List\ \La\n,\x\Ra\Ra\\
\textsf{
Full not protected. List constructors are not protected in their occurrences in function $\Map$.}\\
\end{array}
$

\noindent\begin{tabular}{|c |c |}
\hline
$
\begin{array}{l}
\tiny\begin{array}{lcll}
\Sigma^\q_{-}
& *2 & : & \ql\ \Int\rightarrow\ql\ \Int\\
& ==0 & : & \qh\ \Int\rightarrow\ql\ \Bool\\
& \pi_1 & : & (\ql\ [\ql\ \Int],\ql\ \Int)\rightarrow\ql\ [\ql\ \Int]\\
& \pi_1 & : & (\ql\ [\ql\ \Int],\ql\ \Int)\rightarrow\ql\ [\ql\ \Int]\\
& [] & : & \ql\ [\ql\ \Int]\\
& \id & : & \qh\ \Int\rightarrow\qu\ \Int\\
& : & : & (\qu\ \Int,\ql\ [\ql\ \Int])\rightarrow\ql\ [\ql\ \Int]\\
& -1 & : & \ql\ \Int\rightarrow\ql\ \Int\\
\end{array}
\end{array}
$
&
$
\begin{array}{l}
\tiny\begin{array}{lcll}
\Sigma^\g_{-}
& *2 & : & \z\ \Int\rightarrow\z\ \Int\\
& ==0 & : & \n\ \Int\rightarrow\lo\ \Bool\\
& \pi_1 & : & (\lo\ [\lo\ \Int],\x\ \Int)\rightarrow\lo\ [\lo\ \Int]\\
& \pi_1 & : & (\lo\ [\lo\ \Int],\n\ \Int)\rightarrow\lo\ [\lo\ \Int]\\
& [] & : & \lo\ [\lo\ \Int]\\
& \id & : & \x\ \Int\rightarrow\lo\ \Int\\
& : & : & (\lo\ \Int,\lo\ [\lo\ \Int])\rightarrow\lo\ [\lo\ \Int]\\
& -1 & : & \n\ \Int\rightarrow\n\ \Int\\
\end{array}
\end{array}
$\\
\hline
\end{tabular}\\

\noindent\begin{tabular}{|c |c |}
\hline
$
\begin{array}{l}

\tiny\begin{array}{lcll}
& +1 & : & \ql\ \Int\rightarrow\ql\ \Int\\
& [] & : & \ql\ [\ql\ \Int]\\
& : & : & (\ql\ \Int,\ql\ [\ql\ \Int])\rightarrow\ql\ [\ql\ \Int]\smallskip\\
\end{array}
\\
\tiny\begin{array}{llll}
\Gamma^\q
& \n & : &\ql\ \Int,\\
& \x & : &\ql\ \Int,\\
& \F & : &\ql\ \Int\rightarrow\ql\ \Int,\\
& \List & : &\La\ql\ \Int,\ql\ \Int\Ra\rightarrow\ql\ [\ql\ \Int],\\
& \Map & : & \La\ql\ \Int\rightarrow\ql\ \Int,\ql\ [\ql\ \Int]\Ra\\
& & & \qquad\qquad\qquad\rightarrow\ql\ [\ql\ \Int]\\
\end{array}

\end{array}
$
&
$
\begin{array}{l}

\tiny\begin{array}{lcll}
& +1 & : & \x\ \Int\rightarrow\x\ \Int\\
& [] & : & \xs\ [\lo\ \Int]\\
& : & : & (\lo\ \Int,\lo\ [\lo\ \Int])\rightarrow\xs\ [\lo\ \Int]\smallskip\\
\end{array}
\\
\tiny\begin{array}{llll}
\Gamma^\g
& \n & : &\n\ \Int,\\
& \x & : &\x\ \Int,\\
& \F & : &\Pi\z:\lo\ \Int.\ \z\ \Int,\\
& \List & : &\Pi\La\n,\x\Ra:\La\n\ \Int,\x\ \Int\Ra.\ \lo\ [\lo\ \Int],\\
& \Map & : & \Pi\La\f,\xs\Ra:\\
& & & \La\Pi\z:\lo\ \Int.\ \z\ \Int,\lo\ [\lo\ \Int]\Ra.\ \xs\ [\lo\ \Int]\\
\end{array}

\end{array}
$\\
\hline
\end{tabular}\\
$\tiny\begin{array}{|ll}
\n\ =\ n,\\ \x\ =\ 0,\\ \F\ =\ \lambda \z .\ \z := (*2)\ \z,\\ \List\ =\ \lambda \La\La\Ra,\La\Ra\Ra .\ \If\ (==0)\ \n \ \Then\ []\ \Else\ (\x:\List\ \La\n := (-1)\ \n,\x := (+1)\ \x\Ra),\\ \Map\ = \ \lambda \La\f,\xs\Ra .\ \Case\  \lo\ \xs\ \Of\ (\xs := [],(\z:\zs)\rightarrow\ \xs := (\F\ \z:\Map\ \La\f,\zs\Ra)),\\
\Map\ \La\f,\List\ \La\n,\x\Ra\Ra\\
\hline
\end{array}
$
$P\ \Gamma^\g\ mapl_1^{\Sigma^\g_{mapl_1}}$

\bigskip

$map_2$
$
\tiny\begin{array}{|ll}
\hline
\ara\ =\ \left\lbrace 0,1,...,n\right\rbrace ,\\ \Fun\ =\ \lambda \x .\ (+1)\ \x,\\ \Map\ = \ \lambda \La\ara,\vi,\n,\z\Ra .\ \If\ (\vi==\n) \ \Then\ \La\ara,\vi,\n,\z\Ra\ \Else\ \Let\ \z \equiv \Fun\ \ara[\vi,\z] \  \In\ \Map\ \La\ara[\vi\rightarrow\z],(+1)\ \vi,\n,\z\Ra,\\
\Map\ \La\ara,0,n,0\Ra\\
\textsf{Protected full.}\\
\end{array}
$

\noindent\begin{tabular}{|c |c |}
\hline
$
\begin{array}{l}
\tiny\begin{array}{lcll}
\Sigma^\q_{-}
& +1 & : & \ql\ \Int\rightarrow\ql\ \Int\\
& == & : & (\qh\ \Int,\qh\ \Int)\rightarrow\ql\ \Bool\\
& \cdot[ \cdot , \cdot ] & : & (\qh\ \Array,\qh\ \Int,\ql\ \Int)\rightarrow\ql\ \Int\\
& \cdot[\cdot\!\leftarrow\!\cdot ] & : & (\ql\ \!\Array,\qh\ \!\Int,\qh\ \!\Int)\!\rightarrow\!\ql\ \!\Array\\
& +1 & : & \ql\ \Int\rightarrow\ql\ \Int\\
& 0 & : & \ql\ \Int\\
& n & : & \ql\ \Int\\
& 0 & : & \ql\ \Int\smallskip\\
\end{array}
\\
\tiny\begin{array}{llll}
\Gamma^\q
& \ara & : &\ql\ \Array,\\
& \Fun & : &\ql\ \Int\rightarrow\ql\ \Int,\\
& \Map & : & \La\ql\ \Array,\ql\ \Int,\ql\ \Int,\ql\ \Int\Ra\\
&  & & \qquad\quad \rightarrow\La\ql\ \Array,\ql\ \Int,\ql\ \Int,\ql\ \Int\Ra\\
\end{array}
\end{array}
$
&
$
\begin{array}{l}
\tiny\begin{array}{lcll}
\Sigma^\g_{-}
& +1 & : & \z\ \Int\rightarrow\z\ \Int\\
& == & : & (\vi\ \Int,\n\ \Int)\rightarrow\lo\ \Bool\\
& \cdot[\cdot, \cdot] & : & (\ara\ \Array,\vi\ \Int,\z\ \Int)\!\rightarrow\!\z\ \Int\\
& \cdot[\cdot\!\leftarrow\!\cdot ] & : & (\ara\ \Array,\vi\ \Int,\z\ \Int)\!\rightarrow\!\ara\ \Array\\
& +1 & : & \vi\ \Int\rightarrow\i\ \Int\\
& 0 & : & \lo\ \Int\\
& n & : & \lo\ \Int\\
& 0 & : & \lo\ \Int\smallskip\\
\end{array}
\\
\tiny\begin{array}{llll}
\Gamma^\g
& \ara & : &\ara\ \Array,\\
& \Fun & : &\Pi\z:\lo\ \Int.\ \z\ \Int,\\
& \Map & : & \Pi\La\ara,\vi,\n,\z\Ra:\\
& & &\qquad\La\ara\ \Array,\lo\ \Int,\lo\ \Int,\lo\ \Int\Ra.\\
& & & \qquad \La\ara\ \Array,\vi\ \Int,\n\ \Int,\z\ \Int\Ra\\
\end{array}
\end{array}
$\\
\hline
\end{tabular}\\ 
$\tiny\begin{array}{|ll}
\ara\ =\ \left\lbrace 0,1,..n\right\rbrace ,\\ \Fun\ =\ \lambda \z .\ \z := (+1)\ \z,\\ \Map\ = \ \lambda \La\La\Ra,\vi,\n,\z\Ra .\ \If\ (\vi==\n) \ \Then\ \La\ara,\vi,\n,\z\Ra\\
\qquad\qquad\qquad\qquad\qquad\qquad\qquad\Else\ \Let\ \z \equiv \Fun\ \z := \ara[\vi,\z] \  \In\ \Map\ \La\ara := \ara[\vi\rightarrow\z],\vi := (+1)\ \vi,\n,\z\Ra,\\
\Map\ \La\ara,0,n,0\Ra\\
\hline
\end{array}
$
$P\ \Gamma^\g\ map_2^{\Sigma^\g_{-}}$

\bigskip

\noindent$insl_2$
$
\tiny\begin{array}{|ll}
\hline
\n\ =\ n,\\ \y\ =\ 0,\\ \List\ =\ \lambda \La\n,\y\Ra .\ \If\ (==0)\ \n \ \Then\ \pi_1(\pi_1([],\n),\y)\ \Else\ (\id(\y):\List\ \La(-1)\ \n,(+1)\ \y\Ra),\\ \Ins\ = \ \lambda \La\x,\xs\Ra .\ \Case\  \qh\ \xs\ \Of\ ([\xs](\x:[]),(\z:\zs)\rightarrow\ \If\ (\x\leq\z) \ \Then\ [\xs](\x:(\z:\zs))\ \Else\ [\xs](\z:\Ins\ \La\x,\zs\Ra)),\\
\Ins\ \La n,\List\ \La\n,\y\Ra\Ra\\
\textsf{Protected full.}\\
\end{array}
$

\noindent\begin{tabular}{|c |c |}
\hline
$
\begin{array}{l}
\tiny\begin{array}{lcll}
 ==0 & : & \qh\ \Int\rightarrow\ql\ \Bool\\
 \pi_1 & : & (\ql\ [\ql\ \Int],\ql\ \Int)\rightarrow\ql\ [\ql\ \Int]\\
 \pi_1 & : & (\ql\ [\ql\ \Int],\ql\ \Int)\rightarrow\ql\ [\ql\ \Int]\\
 \left[\right] & : & \ql\ [\ql\ \Int]\\
 \id & : & \qh\ \Int\rightarrow\qu\ \Int\\
 : & : & (\qu\ \Int,\ql\ [\ql\ \Int])\rightarrow\ql\ [\ql\ \Int]\\
 -1 & : & \ql\ \Int\rightarrow\ql\ \Int\\
 +1 & : & \ql\ \Int\rightarrow\ql\ \Int\\
 \left[:\right] & : & (\ql\ \![\ql\ \Int],\ql\ \Int,\ql\ \![\ql\ \Int])\!\rightarrow\!\ql\ \![\ql\ \Int]\\
 \left[\right] & : & \ql\ [\ql\ \Int]\\
 \leq & : & (\qh\ \Int,\qh\ \Int)\rightarrow\ql\ \Bool\\
 \left[:\right] & : & (\ql\ \![\ql\ \Int],\ql\ \Int,\ql\ \![\ql\ \Int])\!\rightarrow\!\ql\ \![\ql\ \Int]\\
 : & : & (\ql\ \Int,\ql\ [\ql\ \Int])\rightarrow\ql\ [\ql\ \Int]\\
 \left[:\right] & : & (\ql\ \![\ql\ \Int],\ql\ \Int,\ql\ \![\ql\ \Int])\!\rightarrow\!\ql\ \![\ql\ \Int]\\
 n & : & \ql\ \Int\smallskip\\
\end{array}
\\
\tiny\begin{array}{llll}
& \n & : &\ql\ \Int,\\
& \y & : &\ql\ \Int,\\
& \List & : &\La\ql\ \Int,\ql\ \Int\Ra\rightarrow\ql\ [\ql\ \Int],\\
& \Ins & : & \La\ql\ \Int,\ql\ [\ql\ \Int]\Ra\rightarrow\
 \ql\ [\ql\ \Int]\\
\end{array}
\end{array}
$
&
$
\begin{array}{l}
\tiny\begin{array}{lcll}
& ==0 & : & \n\ \Int\rightarrow\lo\ \Bool\\
& \pi_1 & : & (\lo\ [\lo\ \Int],\y\ \Int)\rightarrow\lo\ [\lo\ \Int]\\
& \pi_1 & : & (\lo\ [\lo\ \Int],\n\ \Int)\rightarrow\lo\ [\lo\ \Int]\\
& [] & : & \lo\ [\lo\ \Int]\\
& \id & : & \y\ \Int\rightarrow\lo\ \Int\\
& : & : & (\lo\ \Int,\lo\ [\lo\ \Int])\rightarrow\lo\ [\lo\ \Int]\\
& -1 & : & \n\ \Int\rightarrow\n\ \Int\\
& +1 & : & \y\ \Int\rightarrow\y\ \Int\\
& [:] & : & (\xs\ \![\lo\ \Int],\lo\ \Int,\lo\ \![\lo\ \Int])\!\rightarrow\!\xs\ \![\lo\ \Int]\\
& [] & : & \lo\ [\lo\ \Int]\\
& \leq & : & (\lo\ \Int,\lo\ \Int)\rightarrow\lo\ \Bool\\
& [:] & : & (\xs\ \![\lo\ \Int],\lo\ \Int,\lo\ \![\lo\ \Int])\!\rightarrow\!\xs\ \![\lo\ \Int]\\
& : & : & (\lo\ \Int,\lo\ [\lo\ \Int])\rightarrow\lo\ [\lo\ \Int]\\
& [:] & : & (\xs\ \![\lo\ \Int],\lo\ \Int,\lo\ \![\lo\ \Int])\!\rightarrow\!\xs\ \![\lo\ \Int]\\
& n & : & \lo\ \Int\smallskip\\
\end{array}
\\
\tiny\begin{array}{llll}
& \n & : &\n\ \Int,\\
& \y & : &\y\ \Int,\\
& \List & : &\Pi\La\n,\y\Ra:\La\n\ \Int,\y\ \Int\Ra.\ \lo\ [\lo\ \Int],\\
& \Ins & : & \Pi\La\x,\xs\Ra:\La\lo\ \Int,\lo\ [\lo\ \Int]\Ra.\ \xs\ [\lo\ \Int]\\
\end{array}
\end{array}
$\\
\hline
\end{tabular}\\ 
\noindent$\tiny\begin{array}{|ll}
\n\ =\ n,\\ \y\ =\ 0,\\ \List\ =\ \lambda \La\La\Ra,\La\Ra\Ra .\ \If\ (==0)\ \n \ \Then\ []\ \Else\ (\y:\List\ \La\n := (-1)\ \n,\y := (+1)\ \y\Ra),\\ \Ins\ = \ \lambda \La\x,\xs\Ra .\\
\qquad\quad \Case\  \lo\ \xs\ \Of\ (\xs := [\xs](\x:[]),(\z:\zs)\rightarrow\ \If\ (\x\leq\z) \ \Then\ \xs := (\x:(\z:\zs))\\
\qquad\qquad\qquad\qquad\qquad\qquad\qquad\qquad\qquad\qquad\qquad\qquad\Else\ \xs := (\z:\Ins\ \La\x,\zs\Ra)),\\
\Ins\ \La n,\List\ \La\n,\y\Ra\Ra\\
\hline
\end{array}
$
$P\ \Gamma^\g\ insl_2^{\Sigma^\g_{-}}$

\bigskip

\noindent$mapl_2$
$
\tiny\begin{array}{|ll}
\hline
\n\ =\ n,\\ \y\ =\ 0,\\ \F\ =\ \lambda \z .\ (*2)\ \z,\\ \List\ =\ \lambda \La\n,\x\Ra .\ \If\ (==0)\ \n \ \Then\ \pi_1(\pi_1([],\n),\x)\ \Else\ (\id(\x):\List\ \La(-1)\ \n,(+1)\ \x\Ra),\\ \Map\ = \ \lambda \La\f,\xs\Ra .\ \Case\  \qh\ \xs\ \Of\ (\pi_1([],\xs),(\z:\zs)\rightarrow\ [\xs](\f\ \z:\Map\ \La\f,\zs\Ra)),\\
\Map\ \La\f,\List\ \La\n,\y\Ra\Ra\\
\textsf{Protected full.}\\
\end{array}
$

\noindent\begin{tabular}{|c |c |}
\hline
$
\begin{array}{l}
\tiny\begin{array}{lcll}
 *2 & : & \ql\ \Int\rightarrow\ql\ \Int\\
 ==0 & : & \qh\ \Int\rightarrow\ql\ \Bool\\
 \pi_1 & : & (\ql\ [\ql\ \Int],\ql\ \Int)\rightarrow\ql\ [\ql\ \Int]\\
 \pi_1 & : & (\ql\ [\ql\ \Int],\ql\ \Int)\rightarrow\ql\ [\ql\ \Int]\\
 \left[\right] & : & \ql\ [\ql\ \Int]\\
 \id & : & \qh\ \Int\rightarrow\qu\ \Int\\
 : & : & (\qu\ \Int,\ql\ [\ql\ \Int])\rightarrow\ql\ [\ql\ \Int]\\
 -1 & : & \ql\ \Int\rightarrow\ql\ \Int\\
 +1 & : & \ql\ \Int\rightarrow\ql\ \Int\\
 \pi_1 & : & (\ql\ [\ql\ \Int],\ql\ [\ql\ \Int])\rightarrow\ql\ [\ql\ \Int]\\
\left[\right] & : & \ql\ [\ql\ \Int]\\
\left[:\right] & : & (\ql\ \![\ql\ \Int],\ql\ \Int,\ql\ \![\ql\ \Int])\!\rightarrow\!\ql\ \![\ql\ \Int]\smallskip\\
\end{array}
\\
\tiny\begin{array}{llll}
& \n & : &\ql\ \Int,\\
& \y & : &\ql\ \Int,\\
& \F & : &\ql\ \Int\rightarrow\ql\ \Int,\\
& \List & : & \La\ql\ \Int,\ql\ \Int\Ra\rightarrow\ql\ [\ql\ \Int],\\
& \Map & : & \La\ql\ \Int\rightarrow
\ql\ \Int,\ql\ [\ql\ \Int]\Ra\\
& & &\qquad\qquad \qquad\qquad \rightarrow\ql\ [\ql\ \Int]\\
\end{array}
\end{array}
$
&
$
\begin{array}{l}
\tiny\begin{array}{lcll}
& *2 & : & \z\ \Int\rightarrow\z\ \Int\\
& ==0 & : & \n\ \Int\rightarrow\lo\ \Bool\\
& \pi_1 & : & (\lo\ [\lo\ \Int],\x\ \Int)\rightarrow\lo\ [\lo\ \Int]\\
& \pi_1 & : & (\lo\ [\lo\ \Int],\n\ \Int)\rightarrow\lo\ [\lo\ \Int]\\
& [] & : & \lo\ [\lo\ \Int]\\
& \id & : & \x\ \Int\rightarrow\lo\ \Int\\
& : & : & (\lo\ \Int,\lo\ [\lo\ \Int])\rightarrow\lo\ [\lo\ \Int]\\
& -1 & : & \n\ \Int\rightarrow\n\ \Int\\
& +1 & : & \x\ \Int\rightarrow\x\ \Int\\
& \pi_1 & : & (\lo\ [\lo\ \Int],\xs\ [\lo\ \Int])\rightarrow\xs\ [\lo\ \Int]\\
& [] & : & \lo\ [\lo\ \Int]\\
& [:] & : & (\xs\ \![\lo\ \Int],\lo\ \Int,\lo\ \![\lo\ \Int])\!\rightarrow\!\xs\ \![\lo\ \Int]\smallskip\\
\end{array}
\\
\tiny\begin{array}{llll}
& \n & : &\lo\ \Int,\\
& \x & : &\lo\ \Int,\\
& \F & : &\Pi\z:\lo\ \Int.\ \z\ \Int,\\
& \List & : &\Pi\La\n,\x\Ra:\La\n\ \Int,\x\ \Int\Ra.\ \lo\ [\lo\ \Int],\\
& \Map & : & \Pi\La\f,\xs\Ra:\\
& & & \La\Pi\z:\lo\ \Int.\ \z\ \Int,\lo\ [\lo\ \Int]\Ra.\\
& & &\qquad\qquad  \xs\ [\lo\ \Int]\\
\end{array}
\end{array}
$\\
\hline
\end{tabular}\\ 
\noindent$\tiny\begin{array}{|ll}
\n\ =\ n,\\ \x\ =\ 0,\\ \F\ =\ \lambda \z .\ \z := (*2)\ \z,\\ \List\ =\ \lambda \La\La\Ra,\La\Ra\Ra .\ \If\ (==0)\ \n \ \Then\ []\ \Else\ (\x:\List\ \La\n := (-1)\ \n,\x := (+1)\ \x\Ra),\\ \Map\ = \ \lambda \La\f,\xs\Ra .\ \Case\  \lo\ \xs\ \Of\ (\xs := [],(\z:\zs)\rightarrow\ \xs := (\f\ \z:\Map\ \La\f,\zs\Ra)),\\
\Map\ \La\f,\List\ \La\n,\x\Ra\Ra\\
\hline
\end{array}
$
$P\ \Gamma^\g\ mapl_2^{\Sigma^\g_{-}}$

\bigskip

\noindent$fact_{3\g}$
$
\tiny\begin{array}{|ll}
\hline
\w\ =\ 1,\ \x\ =\ n,\\
\Fact\ = \ \lambda \La\w,\x\Ra .\ \If\ (==0)\ \x \ \Then\ \La\w,\x\Ra\ \Else\ \Fact\ \La(\x*\w),(-1)\ \x\Ra,\\
\Fact\ \La\w,\x\Ra\\
\textsf{
Protected full. Command with global variables (only): does not perform parameter passing.}\\
\textsf{It should be noted that the imperative program consists of a simple cycle of the style $\mathsf{while}\ \arb\ \Do\ \arc$ (see}\\
\textsf{definition of this phrase in the theoretical language Iswim \cite{reynolds}).}\\
\end{array}
$

\noindent\begin{tabular}{|c |c |}
\hline
$
\begin{array}{l}
\tiny\begin{array}{lcll}
\Sigma^\q_{-}
& ==0 & : & \qh\ \Int\rightarrow\ql\ \Bool\\
& * & : & (\qh\ \Int,\ql\ \Int)\rightarrow\ql\ \Int\\
& -1 & : & \ql\ \Int\rightarrow\ql\ \Int\\
\end{array}
\\
\tiny\begin{array}{llll}
\Gamma^\q
& \w & : &\ql\ \Int,\\
& \x & : &\ql\ \Int,\\
& \Fact & : & \La\ql\ \Int,\ql\ \Int\Ra\rightarrow\La\ql\ \Int,\ql\ \Int\Ra\\
\end{array}
\end{array}
$
&
$
\begin{array}{l}
\tiny\begin{array}{lcll}
\Sigma^\g_{-}
& ==0 & : & \x\ \Int\rightarrow\lo\ \Bool\\
& * & : & (\x\ \Int,\w\ \Int)\rightarrow\w\ \Int\\
& -1 & : & \x\ \Int\rightarrow\x\ \Int\smallskip\\
\end{array}
\\
\tiny\begin{array}{llll}
\Gamma^\g
& \w & : &\w\ \Int,\\
& \x & : &\x\ \Int,\\
& \f & : & \Pi\La\w,\x\Ra:\La\w\ \Int,\x\ \Int\Ra.\ \La\w\ \Int,\x\ \Int\Ra\\
\end{array}

\end{array}
$\\
\hline
\end{tabular}\\
\noindent$\tiny\begin{array}{|ll}
\w\ =\ 1,\ \x\ =\ n,\\ \f\ = \ \lambda \La\La\Ra,\La\Ra\Ra .\ \If\ (==0)\ \x \ \Then\ \La\w,\x\Ra\ \Else\ \f\ \La\w := (\x*\w),\x := (-1)\ \x\Ra,\\
\f\ \La\w,\x\Ra\\
\hline
\end{array}
$
$P\ \Gamma^\g\ fact_{3\g}^{\Sigma^\g_{fact_{3\g}}}$

\bigskip

\noindent$fib_{5\g}$
$
\tiny\begin{array}{|ll}
\hline
\x\ =\ n,\ \w\ =\ 1,\ \y\ =\ 1,\ \z\ =\ 1,\\
\Fib\ = \ \lambda \La\x,\w,\y,\z\Ra .\ \If\ (==0)\ \x \ \Then\ \La\x,\w,\y,\z\Ra\\
\qquad\qquad\qquad\qquad\Else\ \Let\ \La\x,\w,\y,\z\Ra \equiv \Fib\ \La(-1)\ \x,\w,\y,\z\Ra \  \In\ \Let\ \z \equiv \pi_2(\z,\w) \  \In\ \La\x,\pi_2(\w,\y),(\z+\y),\z\Ra,\\
\Fib\ \La \x,\w,\y,\z\Ra\\
\textsf{
Protected full. Command with global variables (only): does not perform parameter passing.}\\
\end{array}
$

\noindent\begin{tabular}{|c |c |}
\hline
$
\begin{array}{l}
\tiny\begin{array}{lcll}
\Sigma^\q_{-}
& ==0 & : & \qh\ \Int\rightarrow\ql\ \Bool\\
& -1 & : & \ql\ \Int\rightarrow\ql\ \Int\\
& \pi_2 & : & (\ql\ \Int,\qh\ \Int)\rightarrow\ql\ \Int\\
& \pi_2 & : & (\ql\ \Int,\qh\ \Int)\rightarrow\ql\ \Int\\
& + & : & (\qh\ \Int,\ql\ \Int)\rightarrow\ql\ \Int\\
\end{array}
\\
\tiny\begin{array}{llll}
\Gamma^\q
& \x & : &\ql\ \Int,\\
& \w & : &\ql\ \Int,\\
& \y & : &\ql\ \Int,\\
& \z & : &\ql\ \Int,\\
& \Fib & : & \La\ql\ \Int,\ql\ \Int,\ql\ \Int,\ql\ \Int\Ra\rightarrow\\
& & & \quad\qquad\qquad\La\ql\ \Int,\ql\ \Int,\ql\ \Int,\ql\ \Int\Ra\\
\end{array}
\end{array}
$
&
$
\begin{array}{l}
\tiny\begin{array}{lcll}
\Sigma^\g_{-}
& ==0 & : & \x\ \Int\rightarrow\lo\ \Bool\\
& -1 & : & \x\ \Int\rightarrow\x\ \Int\\
& \pi_2 & : & (\z\ \Int,\w\ \Int)\rightarrow\z\ \Int\\
& \pi_2 & : & (\w\ \Int,\y\ \Int)\rightarrow\w\ \Int\\
& + & : & (\z\ \Int,\y\ \Int)\rightarrow\y\ \Int\smallskip\\
\end{array}
\\
\tiny\begin{array}{llll}
\Gamma^\g
& \x & : &\x\ \Int,\\
& \w & : &\w\ \Int,\\
& \y & : &\y\ \Int,\\
& \z & : &\z\ \Int,\\
& \Fib & : & \Pi\La\x,\w,\y,\z\Ra:(\x\ \!\Int,\w\ \!\Int,\\
& & & \y\ \!\Int,\z\ \!\Int).\La\x\ \!\Int,\w\ \!\Int,\y\ \!\Int,\z\ \!\Int\Ra\\
\end{array}
\end{array}
$\\
\hline
\end{tabular}\\
\noindent$\tiny\begin{array}{|ll}
\x\ =\ 20,\ \w\ =\ 1,\ \y\ =\ 1,\ \z\ =\ 1,\\
\Fib\ = \ \lambda \La\La\Ra,\La\Ra,\La\Ra,\La\Ra\Ra .\\
\qquad\quad\If\ (==0)\ \x \ \Then\ \La\x,\w,\y,\z\Ra\ \Else\ \Fib\ \La\x := (-1)\ \x,\w,\y,\z\Ra;\  \z := \w;\ \La\x,\w := \y,\y := (\z+\y),\z\Ra,\\
\Fib\ \La\x,\w,\y,\z\Ra\\
\hline
\end{array}
$
$P\ \Gamma^\g\ fib_{5\g}^{\Sigma^\g_{-}}$

\bigskip

\noindent$case$
$
\tiny\begin{array}{|ll}
\hline
\z=0,\ \Fun\ =\ \lambda \z .\ (*2)\ \z,\ \G\ = \ \lambda \xs .\ \Case\  \ql\ \xs\ \Of\ ([],(\z:\zs)\rightarrow\ (\Fun\ \z:\zs)),\\
\La\z,\G\ (1:2:[])\Ra\\
\textsf{
Protected with incorrect globalization. It is linearly typed, and would be globally typed if we did not prevent parameter globalization in}\\
\textsf{
the $\Case$ phrase. Note in this example that protection does not guarantee correctness without this restriction.}\\
\end{array}
$

\noindent\begin{tabular}{|c |c |}
\hline
$
\begin{array}{l}

\tiny\begin{array}{lcll}
\Sigma^\q_{case}
& *2 & : & \ql\ \Int\rightarrow\ql\ \Int\\
& [] & : & \ql\ [\ql\ \Int]\\
& : & : & (\ql\ \Int,\ql\ [\ql\ \Int])\rightarrow\ql\ [\ql\ \Int]\\
& 1 & : & \ql\ \Int\\
& : & : & (\ql\ \Int,\ql\ [\ql\ \Int])\rightarrow\ql\ [\ql\ \Int]\\
& 2 & : & \ql\ \Int\\
& : & : & (\ql\ \Int,\ql\ [\ql\ \Int])\rightarrow\ql\ [\ql\ \Int]\\
& [] & : & \ql\ [\ql\ \Int]\smallskip\\
\end{array}
\\
\tiny\begin{array}{llll}
\Gamma^\q
&\z & : & \ql \ \Int\\
& \Fun & : &\ql\ \Int\rightarrow\ql\ \Int,\\
& \G & : & \ql\ [\ql\ \Int]\rightarrow\ql\ [\ql\ \Int]\\
\end{array}
\end{array}
$
&
$
\begin{array}{l}
\tiny\begin{array}{lcll}
\Sigma^\g_{case}
& *2 & : & \z\ \Int\rightarrow\z\ \Int\\
& [] & : & \lo\ [\lo\ \Int]\\
& : & : & (\lo\ \Int,\lo\ [\lo\ \Int])\rightarrow\lo\ [\lo\ \Int]\\
& 1 & : & \lo\ \Int\\
& : & : & (\lo\ \Int,\lo\ [\lo\ \Int])\rightarrow\lo\ [\lo\ \Int]\\
& 2 & : & \lo\ \Int\\
& : & : & (\lo\ \Int,\lo\ [\lo\ \Int])\rightarrow\lo\ [\lo\ \Int]\\
& [] & : & \lo\ [\lo\ \Int]\smallskip\\
\end{array}
\\
\tiny\begin{array}{llll}
\Gamma^\g
& \z & : &\z\ \Int,\\
& \Fun & : &\Pi\z:\z\ \Int.\ \z\ \Int,\\
& \G & : & \Pi\xs:\lo\ [\lo\ \Int].\ \lo\ [\lo\ \Int]\\
\end{array}

\end{array}
$\\
\hline
\end{tabular}\\
\noindent$\tiny\begin{array}{|ll}
\z\ =\ 0,\ \Fun\ =\ \lambda \La\Ra .\ \z := (*2)\ \z,\ \G\ = \ \lambda \xs .\ \Case\  \lo\ \xs\ \Of\ ([],(\ep:\zs)\rightarrow\ (\Fun\ \z:\zs)),\\
\La\z,\G\ (1:2:[])\Ra\\
\hline
\end{array}
$
$P\ \Gamma^\g\ case^{\Sigma^\g_{case}}$


\begin{thebibliography}{}




\bibitem{aiken} Aiken, Alexander, Jeffrey S. Foster, John Kodumal, and Tachio Terauchi. Checking and inferring local non-aliasing. In ACM SIGPLAN Conference on Programming Language Design and Implementation (PLDI), San Diego, California, pages 129-140,
June 2003.

\bibitem{aspinall} Aspinall, D., Hofmann, M. Another Type System for In-Place Update, D. Le Metayer (Ed.): ESOP 2002, LNCS 2305, pp. 36?52, 2002.

\bibitem{chirimar} Chirimar, J.,Gunter,C., Riecke, J., Reference Counting as a Computational Interpretation of Linear Logic, Journal of Functional Programming, 6(2), 1995.

\bibitem{draghicescu} M. Draghicescu and S. Purushothaman. A uniform treatment of order of evaluation and aggregate update. Theoretical Computer Science, 118(2):231?262, September 1993.

\bibitem{dosen} Dosen, Kosta. A historical introduction to substructural logics. In K. Do¨en and P. Schroeder-Heister, editors, Substructural Logics, pages 1?30. Oxford University
Press, 1993.


\bibitem{foster} Foster, Jeffrey S., Tachio Terauchi, and Alex Aiken. Flow-sensitive type qualifiers. In ACM SIGPLAN Conference on Programming Language Design and Implementation
(PLDI), Berlin, Germany, pages 1-12, June 2002.

\bibitem{girard} J.Y. Girard, Linear logic, Theoretical Computer Science 50 (1987) 1-102.

\bibitem{girlaf} J.Y. Girard,  Y. Lafont, Linear Logic and Lazy Computation, in: TAPSOFT ?87, Volume 2, LNCS 250 (Springer-Verlag, Pisa) 52-66.

\bibitem{gramagliawlt} H. Gramaglia, Weak-Linear Types. , arXiv:2402.12108v1 [cs:PL] 19 Feb 2024.

\bibitem{gramagliawlg} H. Gramaglia, .

\bibitem{hofmann} M. Hofmann, A type system for bounded space and functional in-place update, in: G. Smolka (Ed.),Programming Languages and Systems,Lecture Notes in Computer Science,Springer,Berlin,2000,pp. 165?179.



\bibitem{kobayashi} Kobayashi, Naoki.  Quasi-Linear Types, In Proceedings ACM Principles of Programming Languages, pages 29?42, 1999. 40, 50

\bibitem{lafont} Yves Lafont, The linear abstract machine, Theoretical Computer Science, Volume 59, Issues 1?2, July 1988, Pages 157-180

\bibitem{odersky} M. Odersky, Observers for linear types. In B. Krieg-Brückner, edi-
tor, ESOP ?92: 4th European Symposium on Programming, Rennes, France,
Proceedings, pages 390?407. Springer-Verlag, February 1992. Lecture Notes
in Computer Science 582.

\bibitem{schmidt} David A. Schmidt, Detecting global variables in denotational specifications, ACM Transactions on Programming Languages and Systems, Volume 7, Issue 2, April 1985 pp 299?310.

\bibitem{sestoft} Peter Sestoft, Replacing function parameters by global variables, Proceedings of the fourth international conference on Functional programming languages and computer architecture, 1989.


\bibitem{reynolds} John C. Reynolds, Theories of Programming Languages, Cambridge University Press,1998.


\bibitem{smith} Smith, Frederick, David Walker, and Greg Morrisett. Alias types. In European Sym-posium on Programming (ESOP), Berlin, Germany, volume 1782 of Lecture Notes in Computer Science, pages 366-381. Springer-Verlag, April 2000.



\bibitem{shankar} Natarajan Shankar. Static analysis for safe destructive updates in a functional language. In A. Pettorossi, editor, 11th International Workshop on Logic-based Program Synthesis and Transformation (LOPSTR 01), Lecture Notes in Computer Science, pages 1?24. Springer-Verlag, 2002.





\bibitem{wadler} P. Wadler, Linear types can change the world! In IFIP TC 2 Working Conference on Programming Concepts and Methods, Sea of Galilee, Israel, April 1990.
Published as M. Broy and C. Jones, editors, Programming Concepts and Methods, North Holland, 1990.

\bibitem{walker} Walker, David. Subestructural Types Systems, in Advanced Topics in Types and Programming Languages. Benjamin C. Pierce, editor. The MIT Press, Cambridge, Massachusetts London, England 2005.


\end{thebibliography}
\end{document}